\documentclass[11pt]{article}
\usepackage{jheppub}
\usepackage{axodraw}
\usepackage{bbm} 
\allowdisplaybreaks 
\newcommand{\settocdepth}[1]{%
  \addtocontents{toc}{\protect\setcounter{tocdepth}{#1}}}

\newcommand{\la}[1]{\label{#1}}

\newcommand{\eq}{Eq.~}
\newcommand{\se}{Sec.~}
\newcommand{\app}{App.~}
\newcommand{\eqs}{Eqs.~}
\newcommand{\nr}[1]{(\ref{#1})}
\newcommand{\nn}{\nonumber \\}
\renewcommand{\(}{\left(}
\renewcommand{\)}{\right)}
\newcommand{\lb}{\left\{}
\newcommand{\rb}{\right\}}
\newcommand{\lk}{\left[}
\newcommand{\rk}{\right]}

\newcommand{\e}{\epsilon}
\newcommand{\order}[1]{{\cal O}(#1)}

\newcommand{\sumint}[1]{\hbox{$\sum$}\!\!\!\!\!\!\!\int_{#1}}
\newcommand{\sumintp}[1]{\hbox{$\sum^\prime$}\!\!\!\!\!\!\!\!\!\int_{#1}}

\newcommand{\intr}{\int_0^\infty\!\!\!\!\!\!{\rm d}r\,}

\newcommand{\gammaE}{{\gamma_{\small\rm E}}}
\newcommand{\gammaEs}{{\gamma}}
\newcommand{\Nf}{N_{\mathrm{f}}}
\newcommand{\csch}[0]{\text{csch}}

\newcommand{\mE}{m_{\mathrm{E}}}
\newcommand{\intB}{{\cal M}_{0,0}}
\newcommand{\intV}{{\cal M}_{1,0}}
\newcommand{\intMbb}{{\cal M}_{2,-2}}
\renewcommand{\vec}[1]{{\mathbf{#1}}}
\newcommand{\pz}{P_0}
\newcommand{\Pz}{P_0}
\newcommand{\qz}{Q_0}
\newcommand{\rz}{R_0}

\newcommand{\intMNb}{{\cal M}_{N,-2}}
\newcommand{\intMcb}{{\cal M}_{3,-2}}
\newcommand{\intVb}{V_2} 
\newcommand{\cS}{{\cal V}} 
\newcommand{\intx}{\int_0^\infty\!\!\!\!\!\!\!{\rm d}x}
\newcommand{\inty}{\int_0^\infty\!\!\!\!\!\!\!{\rm d}y}
\newcommand{\intz}{\int_0^\infty\!\!\!\!\!\!\!{\rm d}z}
\newcommand{\intxy}{\int_0^x\!\!\!\!\!{\rm d}y}
\newcommand{\vc}{V_{3}}
\newcommand{\vd}{V_{4}}
\newcommand{\ve}{V_{5}}
\newcommand{\vf}{V_{6}}
\newcommand{\vg}{V_{7}}
\newcommand{\VfI}{V_I^{\rm f}}
\newcommand{\VfII}{V_{II}^{\rm f}}
\newcommand{\VfIIh}{\hat{V}_{II}^{\rm f}}

\newcommand{\pic}[1]{\;\parbox[c]{30pt}{\begin{picture}(30,30)(0,0)
\SetWidth{1.0}\SetScale{1.0} #1 \end{picture}}\;}
\newcommand{\picb}[1]{\;\parbox[c]{45pt}{\begin{picture}(45,30)(0,0)
\SetWidth{1.0}\SetScale{1.0} #1 \end{picture}}\;}
\def\scfc{0.8}  
\def\phgt{24}   
\def\pwc{24}    
\def\pwcb{36} 
\newcommand{\PIC}[4]{\;\parbox[c]{#2 pt}{\begin{picture}(#2,#3)(0,0)
\SetWidth{1.0}\SetScale{#4} #1 \end{picture}}\;}
\renewcommand{\pic}[1]{\PIC{#1}{\pwc}{\phgt}{\scfc}}
\renewcommand{\picb}[1]{\PIC{#1}{\pwcb}{\phgt}{\scfc}}
\def\ToptVS(#1,#2,#3){\pic{#1(15,15)(15,0,180) #2(15,15)(15,180,360)%
 #3(30,15)(0,15)}}
\def\ToprVB(#1,#2,#3,#4){\picb{#1(30,15)(15,-120,120) #2(30,15)(15,120,240)%
 #3(15,15)(15,60,300) #4(15,15)(15,-60,60)}}
\def\ToprVV(#1,#2,#3,#4,#5){\!\!\picb{#2(26.25,15)(15,256,76)%
 #3(30,30)(15,30) #1(18.75,15)(15,104,284) #4(15,30)(22.5,0)%
 #5(30,30)(22.5,0)}\!\!}
\def\ToprVMij(#1,#2,#3,#4,#5,#6){\pic{#3(15,15)(15,-30,90)%
 #1(15,15)(15,90,210)%
 #2(15,15)(15,210,330) #5(2,7.5)(15,15) #6(15,15)(15,30) #4(28,7.5)(15,15)%
 \Text(12.5,2)[b]{{$\scriptstyle i$}}\Text(13.5,18)[l]{{$\scriptstyle j$}}}}
\def\TopoSB(#1,#2,#3){\picb{#1(0,15)(7.5,15) #2(22.5,15)(15,0,180)%
 #3(22.5,15)(15,180,360) #1(37.5,15)(45,15)}}
\def\Lsc(#1,#2)(#3,#4){\Line(#1,#2)(#3,#4)}
\def\Asc(#1,#2)(#3,#4,#5){\CArc(#1,#2)(#3,#4,#5)}

\title{A new method for taming tensor sum-integrals}

\preprint{BI-TP 2012/32}

\author{Ioan Ghi\c{s}oiu}
\author{and York Schr\"oder\footnote{Current affiliation: Centro de Ciencias Exactas, Departamento de Ciencias B\'asicas, Universidad del B\'io-B\'io,
Avenida Andr\'es Bello 720, Chill\'an, Chile. {\em E-mail:} yschroder@ubiobio.cl.}}

\affiliation{Faculty of Physics, University of Bielefeld,
  33501 Bielefeld, Germany}

\emailAdd{ighisoiu@physik.uni-bielefeld.de}
\emailAdd{yorks@physik.uni-bielefeld.de}

\abstract{We report on the computation of a class of massless bosonic
three-loop vacuum sum-integrals
which are key building blocks for an evaluation of the Debye screening mass 
in hot QCD. Generalizing known techniques and introducing the concept of
tensor reduction by dimensionality shifts (known to the zero-temperature
community since the work of Tarasov in 1996) 
to finite temperature, we are able to treat hitherto unaccessible cases,
which will allow us to finalize the long-term project of
NNLO Debye mass evaluation.}

\begin{document}
\maketitle
\flushbottom

%
\section{Introduction}
\la{se:intro}

Taking as a motivation the startling imbalance between highly developed
analytic, 
systematic algorithmic 
as well as numeric methods
for multi-loop integrals in zero-temperature field theory 
on the one hand (see, e.g.\ \cite{Chetyrkin:1981qh}), 
and the many unsolved technical challenges one is confronted
with when dealing with phenomenological problems 
in cosmological or heavy-ion-collision related contexts 
that are formulated within finite temperature field theory
on the other hand 
(see, e.g.\ \cite{Arnold:1994ps,Braaten:1995jr,Kajantie:2000iz,DiRenzo:2006nh,Laine:2006cp}),
we continue our line of systematic investigation of the latter class
of problems \cite{Schroder:2008ex,Moller:2010xw,Moeller:2012da,Nishimura:2012ee,Schroder:2012hm,Ghisoiu:2012kn}. 
Once more diving into a complicated and rather technical 
issue, our aim is to exhibit and develop generic tools that 
allow for progress on the thermodynamic (equilibrium) front.
Nevertheless, while in this note we merely focus on a specific
class of multi-loop sum-integral that arise in the thermodynamic
setting, our final result will find a concrete application in
the determination of matching parameters (such as the Debye screening mass) 
in the systematic program of an effective theory treatment of hot QCD
thermodynamics \cite{Ginsparg:1980ef,Braaten:1995cm}.
We relegate the finalization of the long-term project
of determining the Debye screening mass $\mE^2$ of hot Yang-Mills theory 
to NNLO \cite{Moeller:2012da}
to an upcoming publication \cite{debyeMass},
and here focus on the well-separated problem of evaluating
the last missing sum-integral in dimensional regularization,
up to its constant term.
Let us just mention here that, once full 3-loop results for 
$\mE^2$ (interpreted as a matching coefficient between 
4d thermal QCD and 3d electrostatic QCD (EQCD) in the framework
of dimensionally reduced effective 
theories \cite{Ginsparg:1980ef,Braaten:1995cm}) 
are available, a certain
contribution of order $g^7$ to the hot QCD pressure can be
deduced, and refer to \se6 of Ref.~\cite{Moeller:2012da} for a more 
detailed discussion of this potential phenomenological application.

There are a number of well-established techniques
that prove useful when evaluating sum-integrals at higher loop order. 
Starting from the seminal work of Arnold and Zhai \cite{Arnold:1994ps}, 
a number of specific cases have appeared in several applications,
ranging from 3-loop sum-integrals needed for 
QCD thermodynamics \cite{Arnold:1994ps,Braaten:1995jr},
to even some 4-loop cases that have served to quantify
high-order corrections in the thermodynamics of scalar
theory \cite{Gynther:2007bw,Andersen:2009ct} 
as well as large $\Nf$ QCD \cite{Gynther:2009qf}.
In each case, the corresponding sum-integrals that appear in the
calculations have been 
evaluated to sufficient depth in their $\e$\/-expansions,
building a (small but valuable) database of master sum-integrals.
There is an urgent need to enlarge this database,
e.g.\ in order to make progress on the precision of
matching computations \cite{Moeller:2012da} (involving moments
of 3-loop self-energies)
or on the determination of the {\em physical leading order} 
(for a justification of this term see e.g.\ \se6 of \cite{Moeller:2012da})
of hot QCD observables \cite{Kajantie:2000iz,Nishimura:2012ee} 
(where further 4-loop masters are required).

More recently, in a series of works we have re-examined
these computational techniques as well as most of the 
known specific cases of sum-integrals, and put them on a more generic
footing, using notation that allows for generalizations 
\cite{Moller:2010xw,Moeller:2012da,Schroder:2012hm,Ghisoiu:2012kn}.
Performing those generalizations for computing previously
unknown cases \cite{Ghisoiu:2012kn}, we have pointed out that
integration-by-parts (IBP) relations can be used profitably here as well.
IBP relations have already been successfully applied in the thermal
context to the reduction process 
itself \cite{Moeller:2012da,Nishimura:2012ee}, 
but now they can also be employed to help
in evaluating the corresponding master sum-integrals that remain after
the reduction algorithm has halted -- mainly in order to
transform infrared divergent parts of dimensionally regularized  
masters in terms of convergent ones \cite{Ghisoiu:2012kn}.
We will see that also in the present context, they play an important role.

In this note, we wish to infuse another new idea to
the art of evaluating sum-integrals, namely to use the ideas
of Tarasov \cite{Tarasov:1996br} for tensor reduction. The main goal is
to avoid the traditional approach of projection methods that -- if
applied to thermal field theory where the rest frame
of the heat bath breaks rotational invariance and introduces
a vector $U=(1,\vec(0))$ that all tensors can depend on -- lead
to inverse structures (such as $1/\vec{p}^2$) 
of a different form than propagators,
hence leading outside the class of sum-integrals that one
has started with (in fact even outside its natural generalization
as suggested by IBP \cite{Schroder:2012hm,Nishimura:2012ee}).
The price to pay is a shift (in units of two) of the
dimensionality of the integral measure, and an increase of powers
of the propagators that are present in the corresponding 
sum-integral \cite{Tarasov:1996br}.
We will argue that this price is not too high,
and show explicitly that one can deal with these 
dimensionally shifted cases.

As a concrete example on which to test our new methods, we
will consider a specific massless bosonic three-loop sum-integral 
of mass dimension two (which we shall name $\intMcb$) here, 
which constitutes one of the last 
building blocks needed for evaluating the Debye screening mass
of thermal Yang-Mills theory.
In fact, in order to stress the generality of our method,
we will treat a more general (infinite) class of sum-integrals 
($\intMNb$, for integer values of $N$) for
most part of the paper, of which $\intMcb$ is just a special case.

The paper is organized as follows.
In \se\ref{se:decomposition}, we introduce the class of
3-loop tensor sum integrals $\intMNb$ on which we wish to test our new idea of
tensor reduction, and explain its decomposition into
scalar pieces. Owing to the structure of the latter (they
all contain two one-loop sub-integrals and are hence of
so-called spectacles-type) we further
decompose them
into pieces that either allow for analytic solutions, or 
are finite such that they can be evaluated numerically.
Sections \ref{se:nonzero} and \ref{se:zero} then deal with
the evaluation of non-zero (Matsubara) modes and zero-modes
in turn, specializing to the specific sum-integral $\intMcb$
for concrete results, which are then summarized in \se\ref{se:result}.
We conclude in \se\ref{se:conclu} and have relegated some 
technical material to the appendices.

%
\section{Decomposition of $\intMNb$}
\la{se:decomposition}

The sum-integrals $\intMNb$ that we shall be concerned with here
can be represented as a subset of a more general class 
as defined by (see \eqs(A.23), (A.30) of Ref.~\cite{Braaten:1995jr} 
and the review \cite{Schroder:2012hm})
\begin{align}
\la{eq:Mclass}
\ToprVMij(\Asc,\Asc,\Asc,\Lsc,\Lsc,\Lsc) \equiv 
{\cal M}_{i,j} &\equiv \sumint{PQR} \frac{\lk(Q-R)^2\rk^{-j}}{\lk P^2\rk^i}\, 
\frac1{Q^2\,(Q+P)^2}\,\frac1{R^2\,(R+P)^2}
\;.
\end{align}
We employ (Euclidean, bosonic) 
four-momenta $P=(\pz,\vec{p})=(2\pi n_p T,\vec{p})$ with 
$P^2 = \pz^2 + \vec{p}^{2}$; the temperature of the thermal 
system is $T$; and the sum-integral symbol stands for
\begin{align}
\sumint{P} \equiv
T\sum_{n_p\in\mathbbm{Z}}\int\frac{\mathrm{d}^d\vec p}{(2\pi)^d}\;,
\quad\mbox{with~~}d=3-2\e \;.
\end{align}
The known non-trivial instances of the class \eq\nr{eq:Mclass} are 
$\intB$ and $\intMbb$ \cite{Arnold:1994ps} of (mass-) dimension four, 
as well as the dimension two case $\intV$ \cite{Andersen:2008bz},
all of which have been re-evaluated in unified notation
in \cite{Schroder:2012hm}.
Furthermore, as has already been pointed out in the latter reference,
the numerators of ${\cal M}_{N,-1}$ can be eliminated
by using the denominator's invariance under momentum shifts
$Q\rightarrow-P-Q$ and $R\rightarrow-P-R$, resulting in 
\begin{align}
{\cal M}_{N,-1} &= 
2\,I_1\,\sumint{PQ}\frac{1}{[P^2]^N\,Q^2\,(P+Q)^2}
-\frac12\,{\cal M}_{N-1,0}\;,
\end{align}
where the first term on the right-hand side (rhs)
has factorized into a trivial 1-loop tadpole $I_1$ (cf.\ \eq\nr{eq:I})
times a two-loop sum-integral of sunset-type 
which can be further factorized into 
a product of one-loop cases by IBP (see e.g.\ \app\ref{se:2loopIBP})
and is hence trivial as well,
while the second term on the rhs is again in the class of \eq\nr{eq:Mclass},
although without scalar products in the numerator and hence much simpler.

The massless bosonic 3-loop vacuum sum-integral $\intMNb$ is 
therefore defined as 
\begin{align}
\la{eq:MNb}
\intMNb &\equiv \sumint{PQR} 
\frac{[(Q-R)^2]^2}{\lk P^2\rk^N Q^2 R^2 (P+Q)^2 (P+R)^2}
\;.
\end{align}
As mentioned above, one representative of this class, $\intMbb$, is
already known \cite{Arnold:1994ps,Braaten:1995jr}, since it has entered 
the 3-loop result for the thermodynamic pressure of hot QCD.
However, its determination (as reviewed in \cite{Schroder:2012hm}) 
was somewhat contrived, mainly due to difficulties in 
treating its numerator structure -- a problem that has triggered
much of what we report here, and that we will alleviate \se\ref{se:taming} 
below.

%
\subsection{Splitting off trivial parts}

Expanding the numerator of the class of massless bosonic 3-loop vacuum 
sum-integrals $\intMNb$ in \eq\nr{eq:MNb} as 
$[(Q-R)^2]^2=4(QR)^2-4(QR)(Q^2+R^2)+Q^4+R^4+2Q^2R^2$,
its spectacles-type structure becomes explicit when using the 
\{scalar,vector,tensor\}-like 2-point functions
\begin{align}
\lb\Pi_{ab}(P),\Pi_{ab}^{\mu}(P),\Pi_{ab}^{\mu\nu}(P)\rb\equiv
\sumint{Q}\frac{\lb 1,Q^\mu,Q^\mu Q^\nu\rb}{[Q^2]^a[(P+Q)^2]^b}
\;,
\end{align}
such that (omitting the argument ($P$) of all functions $\Pi$ 
for brevity from now on)
\begin{align}
\intMNb &= \sumint{P}\frac1{[P^2]^N}\lb
4\Pi_{11}^{\mu\nu}\Pi_{11}^{\mu\nu}
-4\Pi_{01}^{\mu}\Pi_{11}^{\mu}
-4\Pi_{11}^{\mu}\Pi_{01}^{\mu}
+\Pi_{-11}\Pi_{11}
+\Pi_{11}\Pi_{-11}
+2\Pi_{01}\Pi_{01}
\rb\nn
&= \sumint{P}\frac1{[P^2]^N}\lb
4\Pi_{11}^{\mu\nu}\Pi_{11}^{\mu\nu}
-2P^2\Pi_{11}\Pi_{10}
+2\Pi_{10}\Pi_{10}
\rb\\
\la{eq:decomposed}
&= 
4\sumint{P}\frac1{[P^2]^N}\,\Pi_{11}^{\mu\nu}\Pi_{11}^{\mu\nu}
-2I_1^0\sumint{P}\frac{\Pi_{11}}{[P^2]^{N-1}}
+2I_N^0 I_1^0 I_1^0
\;,
\end{align}
where in the second and third line we have used some trivial properties of the 
scalar and vector-like 2-point-functions, which derive from the shifts
$Q\rightarrow-P-Q$ or $Q\rightarrow-Q$ and read
\begin{align}
\Pi_{ab}=\Pi_{ba}\;,\;
\Pi_{a0}=I_a^0\;,\;
\Pi_{00}=I_0^0=\sumint{Q}1=0\;,\;
\Pi_{-1a}=P^2I_a^0+I_{a-1}^0\;;\\
\Pi_{ab}^\mu=-P^\mu\Pi_{ba}-\Pi_{ba}^\mu\;,\;
\Pi_{aa}^\mu=-\frac{P^\mu}2\,\Pi_{aa}\;,\;
\Pi_{a0}^\mu=0\;,\;
\Pi_{0a}^\mu=-P^\mu\,I_a^0
\;.
\end{align}
In \eq\nr{eq:decomposed}, the third term is a product of 1-loop 
sum-integrals and hence known analytically, cf.\ \eq\nr{eq:I}, 
the second term is a 2-loop problem and hence equally trivial 
(it factorizes into 1-loop integrals via IBP, see \eq\nr{eq:L211}), 
while the first term needs further treatment and will be the focus of
the next section.

%
\subsection{Taming the tensors}
\la{se:taming}

To proceed, rewrite 
$\Pi_{11}^{\mu\nu}\Pi_{11}^{\mu\nu}=
\Pi_{11}^{00}\Pi_{11}^{00}
+2\Pi_{11}^{0i}\Pi_{11}^{0i}
+\Pi_{11}^{ij}\Pi_{11}^{ij}$. 
Noting that standard tensor reduction of the 3-vectors present in the
numerator of e.g.\ 
$\Pi_{11}^{0i}=\frac{p^0p^i}{2\vec p^2}\lb I_1^0+\frac{P^2}2\,\Pi_{11}
-2\Pi_{11}^{00}\rb$ 
produces inverses such as $1/\vec{p}^2$,
which we want to avoid since they are not contained in the structure of the
original sum-integrals we started with,
we choose to employ Tarasov's $T$\/-operator technique \cite{Tarasov:1996br}
to trade scalar products 
of 3-vectors for higher dimensions.

To this end, we regard the (massless, 4d) sum-integrals as sums
over corresponding (massive, 3d) spatial integrals, taken at
specific values of the masses. 
In the spirit of Ref.~\cite{Tarasov:1996br}, 
it is then advantageous to generate the irreducible
scalar products of 3-vectors ($\vec q\vec r$ in our case) by derivatives 
acting on a generating function $\exp(-2\alpha\vec q\vec r)$:
\begin{align}
\la{eq:T1}
\sumint{P}\frac1{[P^2]^N}\, \Pi_{11}^{0i}\Pi_{11}^{0i} &=
T^3\!\!\!\sum_{\pz \qz \rz} \!\!\!\left.
\qz\, \rz\, \partial_{-2\alpha}\,
I_{N1111}^{(3-2\e)}(\pz,\qz,\rz,\pz+\qz,\pz+\rz;\alpha)
\right|_{\alpha=0}\;,\\
\la{eq:T2}
\sumint{P}\frac1{[P^2]^N}\, \Pi_{11}^{ij}\Pi_{11}^{ij} &=
T^3\!\!\!\sum_{\pz \qz \rz}\!\!\! \left.
\partial_{-2\alpha}^2\,
I_{N1111}^{(3-2\e)}(\pz,\qz,\rz,\pz+\qz,\pz+\rz;\alpha)
\right|_{\alpha=0}\;,
\end{align}
with the generic (massive, 3d) 3-loop vacuum integral
\begin{align*}
I_{\nu_{1...5}}^{(d)}(m_{1...5};\alpha) &\equiv
\int^{(d)}_{\vec{p}\vec{q}\vec{r}}
\frac{e^{-2\alpha(\vec q\vec r)}}
{[\vec p^2\!+\!m_1^2]^{\nu_1}\,
 [\vec q^2\!+\!m_2^2]^{\nu_2}\,
 [\vec r^2\!+\!m_3^2]^{\nu_3}\,
 [(\vec p\!+\!\vec q)^2\!+\!m_4^2]^{\nu_4}\,
 [(\vec p\!+\!\vec r)^2\!+\!m_5^2]^{\nu_5}} \;.
\end{align*}
This integral has another useful representation, whose 
$\alpha$\/-derivatives will become simple:
introducing Feynman parameters for the propagators
$1/A^\nu=\int_0^\infty\!{\rm d}\alpha\, \alpha^{\nu-1}\,
e^{-\alpha\,A}/\Gamma(\nu)$,
completing squares in the exponential, shifting momenta and rescaling them,
it follows that
\begin{align}
\la{eq:alphaRep}
I_{\nu_{1...5}}^{(d)} &=
\lk \int_{\vec p}^{(d)} \!\!\!\!\! e^{-\vec p^2}  \rk^3
\( \prod_{i=1}^5 
\int_0^\infty\!\!\!\!\!\!{\rm d}\alpha_i\,
\frac{\alpha_i^{\nu_i\!-\!1}\,e^{-\alpha_i m_i^2}}{\Gamma(\nu_i)} \)
\Big[ D(\alpha_j)+\alpha 2 \alpha_4\alpha_5
-\alpha^2(\alpha_1\!+\!\alpha_4\!+\!\alpha_5)\Big]^{-d/2}
\;,
\end{align}
where $D(\alpha_j)$ is the graph polynomial \cite{Bogner:2010kv}
\begin{align}
D(\alpha_j) &=
(\alpha_2+\alpha_4)\alpha_3\alpha_5
+(\alpha_3+\alpha_5)\alpha_2\alpha_4
+\alpha_1(\alpha_2+\alpha_4)(\alpha_3+\alpha_5)
\;.
\end{align}
The Gauss-integral in \eq\nr{eq:alphaRep} depends on the chosen
integral measure; in ours, 
it reads $\lk \int\frac{{\rm d}^dp}{(2\pi)^d}\, e^{-\vec p^2}  
\rk^3=(4\pi)^{-3d/2}$.
Now, note that $\alpha$\/-derivatives acting on \eq\nr{eq:alphaRep}
lower the power of the last term by an integer $n$ (which can be 
interpreted as a shift $d\rightarrow d+2n$), while producing additional 
polynomial pre-factors in the Feynman parameters $\alpha_j$.
The latter can then be pulled out of the integral letting 
$\alpha_j\rightarrow\partial_{-m_j^2}$, and finally be absorbed in
positive shifts of the propagator powers $\nu_i$.
We need the two instances ($d=3-2\e$)
\begin{align*}
\left. \partial_{-2\alpha} I^{(d)}_{N1111}\right|_{\alpha=0}
&=\frac{d}2\,\partial_{-m_4^2}\partial_{-m_5^2}
\left.{\cal D}^+ I^{(d)}_{N1111}\right|_{\alpha=0}
\;=\;\frac{d}2
\left.{\cal D}^+ I^{(d)}_{N1122}\right|_{\alpha=0}
\;,\nn
\left. \partial_{-2\alpha}^2 I^{(d)}_{N1111}\right|_{\alpha=0} 
&=
\frac{d}4\(\partial_{-m_1^2}\!+\!\partial_{-m_4^2}\!+\!\partial_{-m_5^2}\)
\left.{\cal D}^+ I^{(d)}_{N1111}\right|_{\alpha=0}
+\tfrac{d(d+2)}4\,\partial_{-m_4^2}^2\partial_{-m_5^2}^2 
\left.{\cal D}^{++} I^{(d)}_{N1111}\right|_{\alpha=0}\nonumber
\nn
&=
\frac{d}4 \left.{\cal D}^+ \(N\,I^{(d)}_{N+1,1111}+I^{(d)}_{N1121}
+I^{(d)}_{N1112}\)\right|_{\alpha=0}
+d(d+2) \left.{\cal D}^{++} I^{(d)}_{N1133}\right|_{\alpha=0}
\;,
\end{align*}
where ${\cal D}^+ I^{(d)}\equiv (4\pi)^3 I^{(d+2)}$ etc.,
such that \eqs\nr{eq:T1}, \nr{eq:T2} finally read
\begin{align}
\sumint{P}\frac1{[P^2]^N}\,\Pi_{11}^{0i}\Pi_{11}^{0i} &=
\frac{d}2\, {\cal D}^+ \sumint{P}\frac{\Pi_{12}^{0}\Pi_{12}^{0}}{[P^2]^N}
\;,\nn
\sumint{P}\frac1{[P^2]^N}\,\Pi_{11}^{ij}\Pi_{11}^{ij} &=
\frac{N\,d}4\, {\cal D}^+ \sumint{P}\frac{\Pi_{11}\Pi_{11}}{[P^2]^{N+1}}
+\frac{d}2\, {\cal D}^+ \sumint{P}\frac{\Pi_{21}\Pi_{11}}{[P^2]^N}
+d(d+2) {\cal D}^{++} \sumint{P}\frac{\Pi_{31}\Pi_{31}}{[P^2]^N}
\;,\nonumber
\end{align}
and we obtain our final representation of \eq\nr{eq:decomposed} as
\begin{align}
\la{eq:finalRep}
\intMNb &= 
4\sumint{P}\frac{\Pi_{11}^{00}\Pi_{11}^{00}}{P^{2N}}
+4d {\cal D}^+ \sumint{P}\frac{\Pi_{12}^{0}\Pi_{12}^{0}}{P^{2N}}
+N\,d {\cal D}^+ \sumint{P}\frac{\Pi_{11}\Pi_{11}}{P^{2N+2}}
+2d {\cal D}^+ \sumint{P}\frac{\Pi_{21}\Pi_{11}}{P^{2N}}
+\nn&+
4d(d+2) {\cal D}^{++} \sumint{P}\frac{\Pi_{31}\Pi_{31}}{P^{2N}}
-2I_1^0\sumint{P}\frac{\Pi_{11}}{P^{2N-2}}
+2I_N^0 I_1^0 I_1^0 \\
&\equiv
4\vc
+4d \vd 
+N\,d \ve + 2d \vf +4d(d+2) \vg 
-2I_1^0\sumint{P}\frac{\Pi_{11}}{P^{2N-2}}
+2I_N^0 I_1^0 I_1^0
\;.\nonumber
\end{align}
It remains to calculate 1/3/1 3-loop spectacles-type sum-integrals 
$\{\vc,\vd,\ve,\vf,\vg\}$
in $d$/$d\!+\!2$/$d\!+\!4$ dimensions. 
Note however that their structure is quite similar, such
that we will be able to employ a quite generic strategy
for their evaluation, as will be explained below.

%
\subsection{Decomposition of spectacles-type sum-integrals}

Any spectacles-type sum-integral with two generic 2-point 
functions $\Pi_1(P)$, $\Pi_2(P)$
can be identically rewritten as
(suppressing indices of $\Pi$; for a full definition, see \eq\nr{eq:Vdef})
\begin{align}
\la{eq:VVV}
V &\equiv \sumint{P}\frac{(\pz)^m}{[P^2]^n}\,\Pi_1\,\Pi_2
= V^{\rm f(inite)} + V^{\rm d(ivergent)} + V^{\rm z(ero)}\;,\\
\la{eq:Vf}
V^{\rm f}&=
\sumintp{P}\frac{(\pz)^m}{[P^2]^n}\, \Big\{
\frac12\,(\Pi_1-\Pi_1^B)(\Pi_2-\Pi_2^B)
+(\Pi_1-\Pi_1^C)(\Pi_2^B-\Pi_2^D)
\Big\} +\Big\{1\leftrightarrow2\Big\}\;,\\
\la{eq:Vd}
V^{\rm d}&=
\sumintp{P}\frac{(\pz)^m}{[P^2]^n}\, \Big\{
\Pi_1\Pi_2^D
-\Pi_1^B\Pi_2^D
+(\Pi_1^C-\Pi_1^B)(\Pi_2^B-\Pi_2^D)
+\frac12\,\Pi_1^B\Pi_2^B
\Big\} +\Big\{1\leftrightarrow2\Big\}\;,\\
\la{eq:Vz}
V^{\rm z}&=
\sumint{P}\frac{\delta_{\pz}\,(\pz)^m}{[P^2]^n}\,\Pi_1\Pi_2
= \delta_m\,\sumint{P}\frac{\delta_{\pz}}{[P^2]^n}\,\Pi_1\Pi_2
\;,
\end{align}
where the $\Pi^B$ ($T=0$ part; leading UV div), 
$\Pi^D$ (integer power of $P^2$ such that $(\Pi^B\!-\!\Pi^D)$ 
is finite as $\e\rightarrow0$)
and $\Pi^C$ (sub-leading UV divergence 
and sometimes further subtractions)
are chosen such that the terms in $V^{\rm f}$ are finite (and could 
be evaluated numerically in coordinate space at $\e=0$); 
those in $V^{\rm d}$ are simple 2-loop (first term) 
and trivial 1-loop structures;
while $V^{\rm z}$ is the zero mode treated in \se\ref{se:zero}.
The subtraction terms $\Pi^{B,C,D}$ are detailed in \se\ref{se:2pt}.

%
\section{Non-zero modes}
\la{se:nonzero}

The goal of this section is 
to compute the $V^{\rm d}$ of \eq\nr{eq:Vd} analytically, 
and to evaluate the $V^{\rm f}$ of \eq\nr{eq:Vf} numerically,
from simple low-dimensional integral representations.

%
\subsection{Generic formulae for 2-point subtraction terms}
\la{se:2pt}

For generic indices, let us specify the subtraction terms as
\begin{align}
\Pi_{ab0}^B &\equiv \int_Q \frac1{[Q^2]^a[(P+Q)^2]^b}
= \frac{g_{00}(a,b,d+1)}{(P^2)^{a+b-(d+1)/2}} 
\quad{\rm with}\quad
g_{00}(a,b,d)=G(a,b,d) \;,\\
\Pi_{ab1}^B &\equiv U_\mu \int_Q \frac{Q_\mu}{[Q^2]^a[(P+Q)^2]^b} 
= U_\mu \lb P_\mu A(P^2)\rb = \pz\,A(P^2) 
\nn&=\frac{\pz \, g_{10}(a,b,d+1)}{(P^2)^{a+b-(d+1)/2}} \\
&\mbox{with}\quad g_{10}(a,b,d)=\frac12\,\Big[G(a,b-1,d)-G(a-1,b,d)-G(a,b,d)\Big] \;,\\
\Pi_{ab2}^B &\equiv U_\mu U_\nu \int_Q \frac{Q_\mu Q_\nu}{[Q^2]^a[(P+Q)^2]^b}
= U_\mu U_\nu \lb g_{\mu\nu}A(P^2)+P_\mu P_\nu B(P^2) \rb
= A(P^2)+\pz^2 B(P^2)
\nn&= \frac{g_{21}(a,b,d+1)}{(P^2)^{a+b-(d+3)/2}}
+\frac{\pz^2 \, g_{20}(a,b,d+1)}{(P^2)^{a+b-(d+1)/2}}\\
&\mbox{with}\quad g_{21}(a,b,d)=\Big[ 2G(a-1,b,d)+2G(a,b-1,d)+2G(a-1,b-1,d)
-\nn&\hphantom{\mbox{with}\quad g_{21}(a,b,d)=}
-G(a,b,d)-G(a-2,b,d)-G(a,b-2,d)\Big]\frac1{4(d-1)}\\
&\mbox{and }\quad g_{20}(a,b,d)=G(a-1,b,d)-d\,g_{21}(a,b,d)\;,\\
\Pi_{ab0}^C &\equiv \Pi_{ab0}^B +\frac{I_a^0}{(P^2)^b} 
+\frac{I_b^0}{(P^2)^a} \;,\qquad
\Pi_{ab1}^C \equiv \Pi_{ab1}^B -\frac{\pz\,I_b^0}{(P^2)^a} \;,\\
\Pi_{ab2}^C &\equiv \Pi_{ab2}^B +\frac{I_a^2}{(P^2)^b} 
+\frac{I_b^2+\pz^2\,I_b^0}{(P^2)^a} \;,\\
\la{eq:PiD}
\Pi_{abc}^D &\equiv \frac{(P^2)^\e}{(\alpha T^2)^\e}\,\Pi_{abc}^B 
\quad\mbox{(note that this is $\propto(P^2)^{\rm integer}$, 
for any $d={\rm odd}-2\e$)}\;,
\end{align}
where the coefficient functions $g_{ij}$ derive from 
4d rotational invariance.
Since efficient computation needs compact notation,
we summarize the various $\Pi^B$ and $\Pi^C$ as given above as
\begin{align}
\la{eq:PiB}
\Pi_{abc}^B &= \sum_{n=0}^{[c/2]}\frac{(\pz)^{c-2n}\,g_{c,n}(a,b,d+1)}
{(P^2)^{a+b-(d+1)/2-n}} \;,\\
\la{eq:PiC}
\Pi_{abc}^C-\Pi_{abc}^B &= \frac{1+(-1)^c}2\,\frac{I_a^c(d)}{(P^2)^b}
+(-1)^c \sum_{n=0}^{[c/2]}{c\choose 2n} I_b^{2n}(d)\,
\frac{(\pz)^{c-2n}}{(P^2)^a}\;,
\end{align}
where $[c/2]=\{c/2,(c-1)/2\}$ for $c=\{{\rm even},{\rm odd}\}$.

%
\subsection{Generic formulae for divergences $V^{\rm d}$ of non-zero modes}

Due to the structure of $\Pi_{abc}^D$, cf.\ \eq\nr{eq:PiD}, 
\eq\nr{eq:Vd} factorizes as
\begin{align}
\la{eq:Vdiv}
V^{\rm d}&=\sumintp{P}\frac{(\pz)^m}{[P^2]^n}\, \Big\{
\Pi_1\frac{(P^2)^\e}{(\alpha_2 T^2)^\e}
+(\Pi_1^C-\Pi_1^B)\Big(1-\frac{(P^2)^\e}{(\alpha_2 T^2)^\e}\Big)
+\frac{\Pi_1^B}2\,\Big(1-\frac{2(P^2)^\e}{(\alpha_2 T^2)^\e}\Big)
\Big\}\Pi_2^B 
+\nn&+
\big\{1\leftrightarrow2\big\}\;,
\end{align}
where the $\alpha_i$ (cf.\ \eq\nr{eq:PiD}) are constants that might be chosen such as to
simplify the finite parts $V^{\rm f}$, and which also serve
to facilitate comparison with other approaches 
(e.g.\ \cite{Schroder:2012hm,Ghisoiu:2012kn,Andersen:2008bz}).

For the generic spectacles-type 3-loop sum-integral
\begin{align}
\la{eq:Vdef}
V(s_{1...5};s_{6...8}) &\equiv 
\sumint{PQR}\frac{(\pz)^{s_6}\;(\qz)^{s_7}\;(\rz)^{s_8}}
{[P^2]^{s_1}[Q^2]^{s_2}[R^2]^{s_3}[(P+Q)^2]^{s_4}[(P+R)^2]^{s_5}}
\nn&=
\sumint{P}\frac{(\pz)^{s_6}}{[P^2]^{s_1}}\,
\Pi_{s_2s_4s_7}(P)\, \Pi_{s_3s_5s_8}(P)
\end{align}
we therefore write a general expression for the divergent parts 
of their non-zero modes:
\begin{align}
\la{eq:VdRes}
V^{\rm d}(s_{1...8}) &= 
\sum_{n=0}^{[s_8/2]} g_{s_8,n}(s_3,s_5,d+1)\,
f_{\alpha_2}\big(s_{135}-\tfrac{d+1}2-n,s_2,s_4,s_{68}-2n,s_7;d\big)
+\nn&+
\sum_{n=0}^{[s_7/2]} g_{s_7,n}(s_2,s_4,d+1)\,
f_{\alpha_1}\big(s_{124}-\tfrac{d+1}2-n,s_3,s_5,s_{67}-2n,s_8;d\big)\;,
\end{align}
($\alpha_i$\/-independence follows when adding $\VfII$, 
cf.\ \eq\nr{eq:VfIIh})
where the functions $f$ are given by
\begin{align}
f_{\alpha}\big(s_{1...5};d\big) &\equiv \sumintp{P} 
\frac{(\pz)^{s_4}}{(P^2)^{s_1}}\, 
\Big\{\mbox{see \eq\nr{eq:Vdiv}}\Big\}_{s_{235}}
\nn&=
\la{eq:fexpr}
\frac1{(\alpha T^2)^\e}\,L^\prime(s_1-\e,s_{2...5},d) 
+I_{s_2}^{s_5}(d)\,\hat I_{s_{13}}^{s_4}(d,\alpha)
+\nn&+
\sum_{n=0}^{[s_5/2]} \Big[
(-1)^{s_5} {s_5\choose 2n} I_{s_3}^{2n}(d)\,\hat I_{s_{12}}^{s_{45}-2n}(d,\alpha)
+\nn&\qquad\qquad
+\frac12\, g_{s_5,n}(s_2,s_3,d+1)\,\tilde I_{s_{123}-(d+1)/2-n}^{s_4+s_5-2n}(d,\alpha)
\Big]
\end{align}
with the one- and two-loop structures
\begin{align}
\hat I_a^b(d,\alpha) &\equiv I_a^b(d)-(\alpha T^2)^{-\e}I_{a-\e}^b(d)\;,\\
\tilde I_a^b(d,\alpha) &\equiv I_a^b(d)-2(\alpha T^2)^{-\e}I_{a-\e}^b(d)\;,\\
L^\prime(s_{1...5},d)&\equiv L(s_{1...5},d)-\delta_{s_4}A(s_{1...3},s_5,d)\;.
\end{align}
The latter reduces to 1-loop sum-integrals as $L\rightarrow I\cdot I$, as
can be derived systematically via IBP; the cases relevant for the present 
computation are given in appendix \ref{se:2loopIBP}.

%
\subsection{Specific results for divergences $V^{\rm d}$ of non-zero modes}
\la{se:specificVd}

\mbox{}From \eq\nr{eq:VdRes}, we
obtain the desired expansions 
for the divergent pieces of $\{\vc,\vd,\ve,\vf,\vg\}$
needed for the decomposition \eq\nr{eq:finalRep}
of the specific sum-integral $\intMcb$
around $d=3-2\e$,
\begin{alignat*}{2}
\vc^{\rm d}&=V^{\rm d}(31111;022) &\approx& \frac{T^2}{(4\pi)^4}\,
\frac{(4\pi T^2)^{-3\e}}{\e^2}\,\frac1{288}\,
\Big[1
+\(\frac{73}{12}+\gammaE+24\ln G\)\e+\order{\e^2}\Big]\;,\\
\vd^{\rm d}&={\cal D}^+ V^{\rm d}(31122;011) &\approx& \frac{T^{2}}{(4\pi)^4}\,
\frac{(4\pi T^2)^{-3\e}}{\e^2}\,
\Big[0-\frac1{648}\,\e+\order{\e^2}\Big]\;,\\
\ve^{\rm d}&={\cal D}^+ V^{\rm d}(41111) &\approx& \frac{T^{2}}{(4\pi)^4}\,
\frac{(4\pi T^2)^{-3\e}}{\e^2}\,\frac1{432}\,
\Big[1+\(\frac{34}{3}-11\gammaE-120\ln G+24\ln(2\pi)\)\e
\Big]\;,\\
\vf^{\rm d}&={\cal D}^+ V^{\rm d}(32111) &\approx& \frac{T^{2}}{(4\pi)^4}\,
\frac{(4\pi T^2)^{-3\e}}{\e^2}\,\frac{-1}{96}\,
\Big[1+\(\frac{139}{18}-\frac{17}3\,\gammaE-56\ln G+\frac{40}3\ln(2\pi)\)\e
\Big]\;,\\
\vg^{\rm d}&={\cal D}^{++} V^{\rm d}(33311) &\approx& \frac{T^{2}}{(4\pi)^4}\,
\frac{(4\pi T^2)^{-3\e}}{\e^2}\,\frac1{432}\,
\Big[1+\(\frac{26}{5}-\frac{13}5\,\gammaE-\frac{96}5\ln G+\frac{36}5\ln(2\pi)
\)\e\Big]\;,
\end{alignat*}
where $V^{\rm d}(41111)\equiv V^{\rm d}(41111,000)$ etc., 
the Glaisher constant $G$ appears inside a logarithm, with
$12 \ln(G) = 1+\zeta'(-1)/\zeta(-1)$, and we refrain from listing the
somewhat lengthy constant pieces, which contain the dependence on
the arbitrary constants $\alpha_i$ 
as defined in \eq\nr{eq:Vdiv} as well.
As mentioned above, this dependence will precisely cancel that
in $V^{\rm f}$, which is contained in $\VfII$, cf.\ \eq\nr{eq:VfIIh}.

%
\subsection{Finite parts $V^{\rm f}$ of non-zero modes}

With the structure of the subtracted terms as in \eq\nr{eq:Vf}, 
they contribute to $\intMcb$ as
\begin{align}
\la{eq:Mcbfnz}
\intMcb^{\rm nz,f} &= 
4\vc^{\rm f}(31111,022)
+12{\cal D}^+ \vd^{\rm f}(31122,011)
+9{\cal D}^+ \ve^{\rm f}(41111)
+\nn&+
6{\cal D}^+ \vf^{\rm f}(32111)
+60{\cal D}^{++}\vg^{\rm f}(33311)
+\order{\e}
\nn
&\approx \frac{T^2}{(4\pi)^4}\lk n_1+\order{\e}\rk
\;,
\qquad n_1\approx +0.0645513(1)\;,
\end{align}
where we discuss the numerical evaluation of $n_1$,
for which coordinate-space methods prove useful, 
in \app\ref{se:NZfin}.
The value given above was obtained setting all
$\alpha_i=16\pi^2\,e^{3/2-\gammaEs}$.

%
\section{Zero modes}
\la{se:zero}

Here, we treat the $\pz=0$ modes of all 3-loop 
sum-integrals of \eq\nr{eq:finalRep} that are needed for the
specific case $\intMcb$, while discussing the generic strategy. 
In terms of
\begin{align}
\la{eq:Sdef}
S(s_{1...5};s_6,s_7) &\equiv
\sumint{P} \frac{\delta_{\pz}}{[P^2]^{s_1}}\,\Pi_{s_2s_4s_6}\Pi_{s_3s_5s_7}
\quad,\quad
\Pi_{abc}\equiv\sumint{Q}\frac{\qz^c}{[Q^2]^a[(P+Q)^2]^b}
\;,
\end{align}
and using the IBP relations of \app\ref{se:3loopIBP}
to improve their IR behavior, they read
\begin{alignat}{2}
\la{eq:V3zdef}
\vc^{\rm z} &= S(31111;22)&\;=\;& 
\frac{2}{d\!-\!8}\,\Big[ 
\frac{S(12121;22)\!+\!S(12211;22)\!-\!I_2^2 A(221;2,d)}{(d-5)}
\!-\!I_2^2 A(311;2,d)\Big]\;,\\
\la{eq:V4zdef}
\vd^{\rm z} &= {\cal D}^+ S(31122;11) &\;=\;& 0 \quad 
\mbox{after summation in $\delta_{\pz}\Pi_{ab1}$}\;,\\
\ve^{\rm z} &= 
{\cal D}^+ S(41111)&\;=\;& 
\frac{2}{d-8}\,\Big[\vf^z-{\cal D}^+\big(I_2^0 A(411;0,d)\big)\Big]
\;,\\
\vf^{\rm z} &= {\cal D}^+ S(32111)&\;=\;& 
\frac{1}{d\!-\!5}\,\Big[
{\cal D}^+S(22121)
+\tfrac{2}{d-2}{\cal D}^+S(12221)
-{\cal D}^+\big(I_2^0 A(321;0,d)\big)\Big]\;,\\
\la{eq:V11zdef}
\vg^{\rm z} &= {\cal D}^{++} S(33311)&\;=\;& 
2\,\frac{(d^2-13d+38){\cal D}^{++}S(23321)
-(d-10){\cal D}^{++}S(23222)}{(d-2)(d-5)(d-8)}
\;.
\end{alignat}
A generic decomposition into finite part (first line) and remainder 
of zero-mode integrals $S$ is
\begin{align}
\la{eq:zeroModeDec}
S(s_{1...5};s_6,s_7) &= 
\sumint{P} \frac{\delta_{\pz}}{[P^2]^{s_1}} 
\lk\Pi_{s_2s_4s_6}-\Pi_{s_2s_4s_6}^A\rk
\lk\Pi_{s_3s_5s_7}-\Pi_{s_3s_5s_7}^A\rk
+\nn&+
 \sumint{P} \frac{\delta_{\pz}}{[P^2]^{s_1}}\, \Pi_{s_2s_4s_6} \Pi_{s_3s_5s_7}^A
+\sumint{P} \frac{\delta_{\pz}}{[P^2]^{s_1}}\, \Pi_{s_2s_4s_6}^A \Pi_{s_3s_5s_7}
-\nn&-
\sumint{P} \frac{\delta_{\pz}}{[P^2]^{s_1}}\, \Pi_{s_2s_4s_6}^A \Pi_{s_3s_5s_7}^A
\equiv S^{\rm f(inite)}+S^{\rm d(ivergent)}
\;,
\end{align}
where the subtraction terms are defined as
\begin{align}
\la{eq:PiA}
\Pi_{abc}^A &\equiv 
\theta(d\!+\!2\!-\!2a\!-\!2b\!+\!c)\,
\delta_{\pz}\Pi_{abc}^B 
+\theta(2a\!+\!2b\!-\!c\!-\!2\!-\!d)\,
\delta_{\pz}\sumint{Q}\frac{\delta_{\qz}\,\qz^c}{[Q^2]^a[(P-Q)^2]^b}
\nn&=
\theta(d\!+\!2\!-\!2a\!-\!2b\!+\!c)\,
\delta_{\pz}\Pi_{abc}^B 
+\theta(2a\!+\!2b\!-\!c\!-\!2\!-\!d)\,
\delta_{\pz}\delta_c\frac{T\,G(a,b,d)}{(P^2)^{a+b-d/2}}
\;,
\end{align}
which represents 
subtraction of the usual leading UV divergence $\Pi^B$
given by the $T=0$ piece, whose analytical representation was
given in \eq\nr{eq:PiB},
as well as subtraction of 
the zero-mode inside two-point functions $\Pi_{ab0}$.

%
\subsection{Generic results for divergences $S^{\rm d}$ of zero modes}

Noting that the subtraction terms $\Pi_{abc}^A$ are proportional to 
powers of $\vec p^2$, the last term of \eq\nr{eq:zeroModeDec}
vanishes identically since it is scale-free, while the other two 
can be expressed analytically using the 2-loop function $A$:
\begin{align}
\la{eq:Sd}
S^{\rm d}(s_{1...5};s_6,s_7,d) &= 
a(s_{1...7},d)+a(s_{1325476},d)
+0_{\rm scalefree}\;,\\
a(s_{1...7},d) &=
\theta(d\!+\!2\!-\!2s_{24}\!+\!s_6)\,
e_{s_6}
\,g_{s_6,s_6/2}(s_2,s_4,d+1)
A(s_{124}-\tfrac{d+1+s_6}2,s_3,s_5,s_7,d)
+\nn&+
\theta(2s_{24}\!-\!s_6\!-\!2\!-\!d)\,
\delta_{s_6} T\,G(s_2,s_4,d)\,A(s_{124}-\tfrac{d}2,s_3,s_5,s_7,d)\;,
\end{align}
where $e_s\equiv (1+(-1)^s)/2$ is 1(0) for even(odd) index.

%
\subsection{Specific results for divergences $V^{\rm z,d}$ of zero modes}

Collecting and expanding around $d=3-2\e$ dimensions, we finally get 
for the divergent pieces of the zero modes
\begin{align}
\la{eq:V3zexp}
\vc^{\rm z,d} &\approx \frac{T^2}{(4\pi)^4}\,\frac{(4\pi T^2)^{-3\e}}{\e^2}\,
\frac{-1}{48}
\lk 0+1\e+\(\frac{49}{15}+\frac{\pi^2}{20}-3\gammaE+6\ln(2\pi)\)\e^2
+\order{\e^3}\rk\;,\\
\la{eq:V4zexp}
\vd^{\rm z,d} &= 0\;,\\
\la{eq:V15zexp}
\ve^{\rm z,d} &\approx \frac{T^2}{(4\pi)^4}\,\frac{(4\pi T^2)^{-3\e}}{\e^2}\,
\frac{-1}{72} 
\lk 1+\(\frac53-\gammaE+4\ln(2\pi)\)\e+\order{\e^2}\rk\;,\\
\la{eq:V16zexp}
\vf^{\rm z,d} &\approx \frac{T^2}{(4\pi)^4}\,\frac{(4\pi T^2)^{-3\e}}{\e^2}\, 
\frac{5}{144} 
\lk 1+\(\frac{31}{15}-\gammaE+4\ln(2\pi)\)\e+\order{\e^2}\rk\;,\\
\la{eq:V11zexp}
\vg^{\rm z,d} &\approx \frac{T^2}{(4\pi)^4}\,\frac{(4\pi T^2)^{-3\e}}{\e^2}\,
\frac{-1}{240} 
\lk 1+\(\frac{41}{15}-\gammaE+4\ln(2\pi)\)\e+\order{\e^2}\rk\;.
\end{align}
Here, we have of course assumed that all six subtracted 
sum-integrals shown in the first line of \eq\nr{eq:zeroModeDec}
are finite. We prove this assertion by explicit computation
in \app\ref{se:Zfin}, with results given in
\eq\nr{eq:Mfz} below.

%
\subsection{Finite parts $V^{\rm z,f}$ of zero modes}

We now have to write down integral representations 
for the $\Pi^A$\/-subtracted parts of our
zero-modes and
solve them numerically, to prove that they are finite indeed.
The first line of \eq\nr{eq:zeroModeDec}, 
to be evaluated numerically at $d=3$,
contributes to $\intMcb$ as
\begin{align}
\la{eq:Mfz}
\intMcb^{\rm z,f} &= 
\frac45\, \sumint{P}\frac{\delta_{\pz}}{P^2}\,
\Big\{
\lk\Pi_{222}-\Pi_{222}^B\rk\lk\Pi_{112}-\Pi_{112}^B\rk
+\lk\Pi_{212}-\Pi_{212}^B\rk^2
\Big\}
-\nn&-
\frac65\,
{\cal D}^+ \sumint{P}\frac{\delta_{\pz}}{[P^2]^2}\, 
\Big\{
\lk\Pi_{220}-\Pi_{220}^A\rk\lk\Pi_{110}-\Pi_{110}^A\rk
+2P^2\lk\Pi_{220}-\Pi_{220}^A\rk\lk\Pi_{210}-\Pi_{210}^A\rk
\Big\}
+\nn&+
{\cal D}^{++} \sumint{P}\frac{\delta_{\pz}}{[P^2]^2}\, 
\Big\{
96\lk\Pi_{320}-\Pi_{320}^A\rk\lk\Pi_{310}-\Pi_{310}^A\rk
+84\lk\Pi_{320}-\Pi_{320}^A\rk\lk\Pi_{220}-\Pi_{220}^A\rk
\Big\}
\nn
&\approx \frac{T^2}{(4\pi)^4}\lk n_2+\order{\e}\rk
\;,
\qquad n_2\approx +0.24983747686(1)\;,
\end{align}
where we discuss the 
necessary integral representations as well as their 
numerical evaluation leading to $n_2$ in \app\ref{se:Zfin}.

%
\section{Result for $\intMcb$}
\la{se:result}

According to \eqs\nr{eq:VVV}-\nr{eq:Vz} 
and \eqs\nr{eq:V3zdef}-\nr{eq:zeroModeDec},
each scalar spectacle-type sum-integral is decomposed as
$V=V^{\rm f}+V^{\rm d}+V^{\rm z,f}+V^{\rm z,d}$,
and according to \eq\nr{eq:finalRep} 
(with \eq\nr{eq:L211} for the 2-loop case therein),
the master integral $\intMcb$ can be written exclusively 
in terms of such scalar spectacle-types and trivial 1-loop integrals.
Collecting and expanding, we finally obtain
\begin{align}\la{eq:final}
\mu^{6\e}\,\intMcb &\approx \frac{T^2}{(4\pi)^4}
\(\frac{\mu^2}{4\pi T^2}\)^{3\e}
\frac1{\e^2}
\lk -\frac{5}{36} 
+\!\(\frac{1}{216}\!-\!\frac{5}{36}\,\gammaE\!-\!\frac{10}3\,\ln G\)\e 
+\!m\,\e^2
+\!\order{\e^3} \rk \,,\\
\mbox{with~~}m&=
\frac{251}{1620}
+\frac{83}{72}\,\gammaE^2
+\gamma_E\(-\frac{559}{1080}+\frac23\ln(2\pi)-\frac23\ln G
  +\frac7{1080}\,\zeta(3)-\frac1{3780}\,\zeta(5)\)
-\nn&
-\frac{757}{2160}\,\pi^2
-\frac23\ln^22
-\frac{53}9\ln G
+\frac7{15}\ln(2\pi)
-\frac23\ln(4\pi)\ln\pi
+\frac{28}9\,\gamma_1
+\nn&
+\frac{37}{3240}\,\zeta(3)
+\frac{113}{226800}\,\zeta(5)
-\frac7{540}\,\zeta'(3)
+\frac1{1890}\zeta'(5)
+2\zeta''(-1)
+n_1+n_2
\\&\approx n_1+n_2-6.1720481392962725701 \approx -5.8576594(1)\;.
\end{align}
where the Glaisher constant $G$ appears inside a logarithm, with
$12 \ln(G) = 1+\zeta'(-1)/\zeta(-1)$,
and various zeta values as well as the Stieltjes constant $\gamma_1$,
defined by $\zeta(1+\e)=1/\e+\gammaE-\gamma_1\e+\order{\e^2}$, 
enter the constant term.

While it might at first sight seem that we have used an unnecessarily
generic notation in our decomposition and treatment of the 
various contributing terms, let us remark that this was  
done in order to provide a setup that allows for solving 
many more sum-integrals than just the special case $\intMcb$ given above.
Indeed, as we quickly demonstrate in \app\ref{se:checks},
the two previously known sum-integrals 
$\intV$ \cite{Andersen:2009ct} and $\intVb$ \cite{Ghisoiu:2012kn}
(which each had required quite some effort to evaluate)
are reproduced by our generic formulae, and there is no doubt
that the same could be said about 
$\intMbb$ \cite{Arnold:1994ps,Braaten:1995jr,Schroder:2012hm},
although we have not explicitly checked that.

%
\section{Conclusions}
\la{se:conclu}

In this work, we have made important progress in two respects. 
First, by transferring proven technology from zero-temperature
field theory to the finite-temperature case 
(the $T$\/-operators of \se\ref{se:taming}),
we were able to map tensor sum-integrals onto scalar ones.
Naturally, although we have tested this method on a specific 
case only, it is applicable much more generally.
Much in the spirit of Ref.~\cite{pirsig}, by using
the new tensor technique we have dissected the problem 
to its clean core, finding a class of spectacles-type
sum-integrals that are amenable to known techniques.

Second, we have worked out generic solutions
for these massless bosonic 3-loop spectacles-type
sum-integrals, which are applicable to a wide class 
of such integrals.
Let us note that this is only the third instance of treating 
a more generic class of sum-integrals in a single computation
(the first two such instance were recorded 
in \cite{Moller:2010xw,Moeller:2012da}),
and represents the first steps beyond the case-by-case 
analyses found in the literature -- a development urgently
needed in order to close the glaring gap between 
well-established generic integration techniques in 
zero-temperature field theory and the few painstakingly 
derived cases that are known at finite temperatures.

A generalization of our results to sum-integrals involving
fermions should be possible (at least for vanishing masses 
and zero chemical potentials) in a straightforward 
manner, and could in fact even turn out to be structurally
simpler due to the absence of zero-modes in fermionic lines,
which had cost us quite some effort in the bosonic case as treated 
in \se\ref{se:zero} above.
There are a number of well-defined applications that
await the evaluation of such fermionic sum-integrals, 
such as the $\Nf\neq0$ parts of EQCD matching coefficients
as listed in \cite{Moeller:2012da}.

As an application, of our generic formulae, we have
given results for the new massless 3-loop sum-integral
$\intMcb$, which represents a key building block
of the bosonic part of the Debye screening mass in 
hot QCD \cite{Moeller:2012da,debyeMass}.
Reassuringly, our generic formalism reproduces the two
previously known sum-integrals of the spectacles-type class
without effort, 
as we have demonstrated in \app\ref{se:checks}.

{\bf Note added:} The single $\e$\/-pole of $\intMcb$ in \eq\nr{eq:final} is now 
consistent with a renormalizability criterion in \cite{Bernardo:2026whs}, 
whose authors we thank for communicating an issue with our earlier result to us. 
It transpired that an erroneous sign of the 2-loop reduction \eq\nr{eq:L312} 
had caused a wrong result for $\vd^{\rm d}$ in \se\ref{se:specificVd}
which in turn led to a wrong rational term of the single $\e$\/-pole of our final 
result for $\intMcb$ (as well as a modification of the rational and the $\ln G$ 
term of its constant part $m$) in an earlier version of the manuscript.

\acknowledgments

Y.S.~thanks Hide and Shoko Goyahso for hospitality
while part of this work was done.
The work of I.G.~is supported by the Deutsche
Forschungsgemeinschaft (DFG) under grant no.~GRK~881.
Y.S.~is supported by the Heisenberg program of the DFG, 
under contract no.~SCHR~993/1. 
We are indebted to Luis Gil and Philipp Schicho for pointing out 
an error in an earlier version of the manuscript, and to Eduardo Navarro
and Emilio Viacava for verifying parts of the calculation.


\begin{appendix} 
\settocdepth{1} 

%
\section{Standard integrals}
\la{se:functions}

For convenience, we collect here some of the basic functions used above, 
as defined in \cite{Moller:2010xw}. 
They are the massless 1-loop propagator at zero temperature 
\begin{align}
\la{eq:G}
G(s_1,s_2,d) &\equiv \(p^2\)^{s_{12}-\frac{d}2}\int\frac{{\rm d}^dq}{(2\pi)^d}
\,\frac1{[q^2]^{s_1}[(q-p)^2]^{s_2}}
=
\frac{\Gamma(\frac{d}2-s_1)\Gamma(\frac{d}2-s_2)\Gamma(s_{12}-\frac{d}2)}
{(4\pi)^{d/2}\Gamma(s_1)\Gamma(s_2)\Gamma(d-s_{12})} \;;
\end{align}
the 1-loop bosonic tadpoles 
\begin{align}
\la{eq:I}
I_s^a \equiv \sumint{Q} \frac{|\qz|^a}{[Q^2]^s} 
= \frac{2T\,\zeta(2s-a-d)}{(2\pi T)^{2s-a-d}}\,
\frac{\Gamma(s-\frac{d}2)}{(4\pi)^{d/2}\Gamma(s)}
\;,\quad
I_s \equiv \sumint{Q} \frac1{[Q^2]^s} 
= I_s^0\;;
\end{align}
a specific 2-loop tadpole
\begin{align}
\la{eq:A}
A(s_1,s_2,s_3;s_4,d) &\equiv \sumint{PQ} 
\frac{\delta_{\qz}|\pz|^{s_4}}{[Q^2]^{s_1}[P^2]^{s_2}[(P-Q)^2]^{s_3}}
\\&=
\frac{2T^2\,\zeta(2s_{123}-2d-s_4)}{(2\pi T)^{2s_{123}-2d-s_4}}\,
\frac{\Gamma(s_{13}-\frac{d}2)\Gamma(s_{12}-\frac{d}2)
\Gamma(\frac{d}2-s_1)\Gamma(s_{123}-d)}
{(4\pi)^d\Gamma(s_2)\Gamma(s_3)\Gamma(d/2)\Gamma(s_{1123}-d)}
\;,\nonumber
\end{align}
where the shorthand $s_{abc...} \equiv s_c+s_b+s_c+...\;$.

%
\section{IBP relations for 2-loop sum-integrals}
\la{se:2loopIBP}

It is known that, for integer indices $s_{1...5}$, 
all 2-loop sum-integrals of the form
\begin{align}
\la{eq:L}
L(s_{1...5},d) &\equiv
\sumint{PQ} \frac{(\pz)^{s_4}\;(\qz)^{s_5}}{[P^2]^{s_1}[Q^2]^{s_2}[(P+Q)^2]^{s_3}}
\end{align}
are trivial in the sense that by systematic use of 
IBP relations \cite{Nishimura:2012ee} they
reduce to products of 1-loop sum-integrals (which are in turn known analytically,
cf.\ \eq\nr{eq:I}).
We read the specific bosonic cases that are needed for the
present calculation from our algorithmically generated tables, 
to get 
\begin{align}
L(331,00,d) &= -\frac{12(d-8)(d-5)}{(d-2)(d-4)(d-9)(d-11)}\,
I_4^0(d) I_3^0(d) \;,\\
L(311,00,d) &= -\frac4{(d - 2)(d - 7)}\, I_3^0(d) I_2^0(d) \;,\\
L(221,00,d) &=0 \;,\\
\la{eq:L312}
L(312,11,d) &= -\frac{(d - 6) (d - 5) (d - 3)}{2(d - 9)(d - 7)(d - 2)}\, 
I_3^0(d) I_2^0(d) \;,\\
\la{eq:L211}
L(211,00,d) &= -\frac1{(d-5)(d-2)}\, I_2^0(d) I_2^0(d)\;,\\
\la{eq:L21102}
L(211,02,d) &= \frac{d - 3}{d - 5}\, I_2^0(d) I_1^0(d)\;,\\
L(311,22,d) &= -\frac{(d - 4) (d^2 - 8 d + 19)}{4(d - 7)(d - 5)}\, 
I_2^0(d) I_1^0(d)\;.
\end{align}
These reductions can be verified by the generic 2-loop factorization formula now available from \cite{Davydychev:2023jto}, taking into account that the latter paper uses a 2-loop momentum family that differs from \eq\nr{eq:L} by a loop momentum shift $Q\to-Q$. Note that of the examples listed above, only the overall sign of \eq\nr{eq:L312} is sensitive to this conversion of conventions. 

%
\section{IBP relations for 3-loop zero modes}
\la{se:3loopIBP}

For the zero-mode sum-integrals $S$ defined in \eq\nr{eq:Sdef},
for which the generic IBP relation  
\begin{align}
0&=\partial_{\vec p}\vec p \circ S(s_{1...5};s_6,s_7)\nn
\la{eq:Sibp}
&=\lb(d-2s_1-s_4-s_5)+s_4{\bf 4}^+({\bf 2}^--{\bf 1}^-)
+s_5{\bf 5}^+({\bf 3}^--{\bf 1}^-)\rb\circ S(s_{1...5};s_6,s_7)
\end{align}
holds and the additional symmetries $\delta_{\pz}\Pi_{ab,{\rm odd}}=0$ and
$\delta_{\pz}\Pi_{abc}=(-1)^c\delta_{\pz}\Pi_{bac}$ can be used,
we give a number of linear relations here that prove useful
in \se\ref{se:zero} of the main text.
Acting with the IBP \eq\nr{eq:Sibp} on the sum
$\frac{d-2n+1}2\, S(n1111;ee)+S(n-1,2111;ee)$ (with $e$ an even integer)
and using symmetries gives the generic relation
\begin{align}
0&=\frac12\,(d-2n+1)(d-2n-2)S(n1111;ee)-S(n-2,2121;ee)-S(n-2,2211;ee)
+\nn&+
(d-2n+1)I_2^e A(n11;e,d)+I_2^e A(n-1,21;e,d)
\nonumber\;,
\end{align}
which when applied at $\lb n,e\rb=\lb 3,2\rb$ 
gives
\begin{align}
0&=\frac12\,(d-5)(d-8)S(31111;22)-S(12121;22)-S(12211;22)
+\nn&+
(d-5)I_2^2A(311;2,d)+I_2^2 A(221;2,d)
\nonumber\;.
\end{align}
In complete analogy, 
acting with the IBP \eq\nr{eq:Sibp} on
$S(41111)$, on the sum
$S(32111)+\frac{1}{d-4}\,S(22211)$,
as well as on the sum
$S(32222)-(d-10) S(33212)+\frac{d^2-19d+86}2\, S(33311)$ 
and using symmetries gives
\begin{align}
0&=(d-10)S(41111)-2S(32111)+2I_2^0 A(411;0,d)
\;,\nn
0&=(d-7)S(32111)-S(22121)-\tfrac{2}{d-4}\,S(12221)+I_2^0 A(321;0,d)
\nonumber\;,\\
0&=\frac12\,(d-6)(d-9)(d-12)S(33311)
+(d-14)S(23222)
-(d^2-21d+106)S(23321)
\nonumber\;.
\end{align}

%
\section{Numerical evaluation of finite integrals}

In this Appendix, we treat the finite parts 
of non-zero (\ref{se:NZfin}) and zero-modes (\ref{se:Zfin})
that are needed in the main text, cf.\ \eqs\nr{eq:Mcbfnz} and \nr{eq:Mfz}.

%
\subsection{Contribution to $\intMcb$ from finite 
parts $V^{\rm f}$ of non-zero modes}
\la{se:NZfin}

Here, we discuss the evaluation of the finite integrals $V^{\rm f}$
as defined in \eq\nr{eq:Vf}, and as needed 
for \eq\nr{eq:Mcbfnz} of the main text.
They can be treated at $\e=0$ in coordinate space. Note, 
however, that we need some of them in shifted dimensions (in fact, for 
$d\in\{3,5,7\})$.
We will first provide the (inverse, spatial) Fourier transforms 
of propagators $1/P^2$ as well as two-point functions $\Pi(P)$,
and then reduce the coordinate-space representation of the $V^{\rm f}$
to simple integrals as far as possible, 
once more in a generic way with unspecified indices.
We will profit from the fact that we are interested in odd $d$,
in which case the Bessel functions coming from the Fourier transforms
reduce to Bessel polynomials.
The specific cases that we need for $\intMcb$ are then approximated
numerically.

The $d$\/-dimensional angular averages ($\rm{Re}(d)>1$) result in 
Bessel functions of first kind
\begin{align}
\int{\rm d}^d\vec{r}\,f(r)\,e^{i\vec{p}\vec{r}}
&= \intr r^{d-1}\,f(r)\,
\frac{2\pi^{\frac{d-1}2}}{\Gamma(\frac{d-1}2)}
\int_{-1}^1\!\!\!\!{\rm d}u\,(1-u^2)^{\frac{d-3}2}\,e^{ipru}\nn
&= \intr r^{d-1}\,f(r)\,
2(2\pi)^{\frac{d-1}2}\,(pr)^{2-d}\,j_{\frac{d}2-1}(pr)\\
\la{eq:jtrig}
&\mbox{with~~}j_n(x)=\sqrt{\frac\pi2}\,x^n\,J_n(x)\;,\\
&\;j_{\{\frac12,\frac32,\frac52\}}=
\lb\sin(x),\sin(x)-x\cos(x),3\sin(x)-3x\cos(x)-x^2\sin(x)\rb\;,\nonumber
\end{align}
such that propagators transform into modified Bessel functions of 2nd kind
($0\!<\!\rm{Re}(d)\!<\!4s\!+\!1$)
\begin{align}
\frac1{(\vec p^2+m^2)^s} &=
\int{\rm d}^d\vec{r}\,e^{i\vec{p}\vec{r}}
\int\frac{{\rm d}^d\vec{k}}{(2\pi)^d}\,\frac{e^{-i\vec{k}\vec{r}}}
{(\vec k^2+m^2)^s} 
=\int{\rm d}^d\vec{r}\,e^{i\vec{p}\vec{r}}
\frac{2\,r^{2s-d}}{(2\pi)^{\frac{d+1}2}}
\int_0^\infty\!\!\!\!\!\!{\rm d}k\,\frac{k\,j_{\frac{d}2-1}(k)}{(k^2+m^2r^2)^s}\nn
&=\int{\rm d}^d\vec{r}\,e^{i\vec{p}\vec{r}}
\frac{2^{1-s}}{(2\pi)^{\frac{d}2}}\,\frac1{\Gamma(s)}
\(\frac{m^2}{r^2}\)^{\frac{d-2s}4}
K_{s-\frac{d}2}(\sqrt{m^2r^2})\nn
&=\int{\rm d}^d\vec{r}\,e^{i\vec{p}\vec{r}}
\frac{2^{-s}}{(2\pi)^{\frac{d-1}2}}\,\frac1{\Gamma(s)}
\(\frac{m^2}{r^2}\)^{\frac{d-2s}4}
\frac{e^{-\sqrt{m^2r^2}}}{(m^2r^2)^{\frac14}}\,
\kappa_{s-\frac{d}2}(\sqrt{m^2r^2})\\
&\mbox{with~~}\kappa_n(x)=\sqrt{\frac{2x}\pi}\,e^x\,K_n(x)\;,\\
&\;\kappa_{\pm\{\frac12,\frac32,\frac52\}}(x)=
\lb1,1+\frac1x,1+\frac3x+\frac3{x^2}\rb\;,
\end{align}
and we get the Fourier transforms\footnote{From \eqs\nr{eq:PiD}, 
\nr{eq:PiB}, \nr{eq:PiC} we see that 
$\hat{f}^X_{d,abc}(x,-n)=(-1)^c\,\hat{f}^X_{d,abc}(x,n)$ for all 
$X\in\{\Pi,B,C,D\}$.}
\begin{align}
\la{eq:FTs}
\Pi^{(d)}_{s_{123}}(P)&\equiv T\sum_{\qz}\int\frac{\rm{d}^d\vec{q}}{(2\pi)^d}\,
\frac{(\qz)^{s_3}}{[Q^2]^{s_1}[(P+Q)^2]^{s_2}}\nn
&=\int{\rm d}^d\vec{r}\,e^{i\vec{p}\vec{r}}\,e^{-|\pz|r}\,
\frac{T^d\,(2\pi T)^{d+1+s_3-2s_{12}}}
{2^{s_{12}}\Gamma(s_1)\Gamma(s_2)(2\pi Tr)^{d+1-s_{12}}}\,
\hat{f}^\Pi_{d,s_{123}}(2\pi Tr,\tfrac{\pz}{2\pi T})\\
\mbox{with}&\;\hat{f}^\Pi_{d,abc}(x,n)\equiv
\sum_j \frac{j^c\,e^{-x(|j|+|j+n|-|n|)}}
{|j|^{a-\frac{d-1}2}|j+n|^{b-\frac{d-1}2}}\,
\kappa_{a-\frac{d}2}(|j|x)\,
\kappa_{b-\frac{d}2}(|j+n|x) \;.
\end{align}

%
\subsubsection{First part $\VfI$ of \eq\nr{eq:Vf}}

A generic form for the first part of $V^{\rm f}$ 
from \eq\nr{eq:Vf} (plus its $\{1\!\leftrightarrow\!2\}$ part) is now
\begin{align}
\VfI(d,s_{1...8}) &\equiv
\sumintp{P} \frac{(\Pz)^{s_6}}{(P^2)^{s_1}}
\(\Pi_{s_{247}}-\Pi_{s_{247}}^B\)
\(\Pi_{s_{358}}-\Pi_{s_{358}}^B\)\\
&=
\frac{T^{2d} (2\pi T)^{d+3+s_{678}-2s_{12345}}}
{2^{s_{12345}-1}\Gamma(s_1)\Gamma(s_2)\Gamma(s_3)\Gamma(s_4)\Gamma(s_5)}\,
{\sum_{n}}'
\intx
\inty\;
e^{-|n|(x+y)}\times\nn
&\times
\frac{n^{s_6}|n|^{4-d-2s_1}}{x^{d-s_{24}}y^{d-s_{35}}}\,
h_{d,s_1}(|n|x,|n|y)\,
\hat{f}_{d,s_{247}}^{\Pi-B}(x,n)\,
\hat{f}_{d,s_{358}}^{\Pi-B}(y,n)\;,\\
h_{d,s}(x,y) &= \frac{2^s\Gamma(s)}\pi
\intz\;
\frac{z^{3-d}}{(1+z^2)^s}\,
j_{\frac{d-2}2}(xz)\,
j_{\frac{d-2}2}(yz)\;,\\
\la{eq:fPCd}
\hat{f}_{d,abc}^{\Pi-B}(x,n) &=\sum_j
\frac{j^c\,e^{-x(|j|+|j+n|-|n|)}}{|j|^{a-\frac{d-1}2}\,|j+n|^{b-\frac{d-1}2}}\,
\kappa_{a-\frac{d}2}(|j|x)\,
\kappa_{b-\frac{d}2}(|j+n|x)
-\nn&-
\frac{n^c}{|n|^c}\,
\sum_{j=0}^{[c/2]}
\frac{\Gamma(a)\Gamma(b)(4\pi)^{\frac{d+1}2}}{\Gamma(a+b-\tfrac{d+1}2-j)}\,
\frac{g_{c,j}(a,b,d+1)\,\kappa_{a+b-d-j-1/2}(|n|x)}{(x/2)^j\,|n|^{j+a+b-c-d}}
\;.
\end{align}
In an expansion around odd $d$ and for integer indices $s_i$, 
the Bessel functions $\kappa$ reduce to (reverse) Bessel polynomials
$\kappa_s(x)=\frac{\theta_{|s|-1/2}(x)}{x^{|s|-1/2}}$,
such that ($\alpha=a+b-d-j-1/2$)
\begin{align}
\la{eq:fPC3d}
\hat{f}_{d,abc}^{\Pi-B}(x,n) &\approx
\sum_{k=0}^{|a-\frac{d}2|-\frac12}
\sum_{\ell=0}^{|b-\frac{d}2|-\frac12}
\frac{(2|a-\frac{d}2|-1-k)!}{(|a-\frac{d}2|-\frac12-k)!k!}\,
\frac{(2|b-\frac{d}2|-1-\ell)!}{(|b-\frac{d}2|-\frac12-\ell)!\ell!}\,
\frac{s_{c,k-|a-\frac{d}2|-a+\frac{d}2,\ell-|b-\frac{d}2|-b+\frac{d}2}(x,n)}
{(2x)^{|a-\frac{d}2|+|b-\frac{d}2|-1-k-\ell}}
-\nn&-
\frac{n^c}{|n|^c}\,
\frac{\Gamma(a)\Gamma(b)}{x^{d+c-a-b}}\,
\sum_{j=0}^{[c/2]}
\frac{(4\pi)^{\frac{d+1}2}\,g_{c,j}(a,b,d+1)}{\Gamma(\alpha+\frac{d}2)}
\sum_{k=0}^{|\alpha|-\frac12}
\frac{(2|\alpha|-1-k)!}{(|\alpha|-\frac12-k)!\,k!}\,
\frac{(|n|x)^{k-\alpha-|\alpha|+c-2j}}{2^{|\alpha|-\frac12-k-j}}
+\order{\e}\\
&\mbox{where~~}s_{cab}(x,n)=\sum_j e^{-x(|j|+|j+n|-|n|)}\,j^c\,|j|^a\,|j+n|^b
= \hat{s}_{cab}(\coth(x),n)\;,\\
&\hat{s}_{cab}(y,-n)=(-1)^c\,\hat{s}_{cab}(y,n)\;,
\;\;\hat{s}_{0ab}(y,\mathbbm{Z})=\hat{s}_{0ba}(y,\mathbbm{Z})\;,
\;\;\hat{s}_{cab}(y,0)=\hat{s}_{c,a\!+\!b,0}(y,0)\;,\nn
&\hat{s}_{c_{\rm even}ab}(y,n)=\hat{s}_{0,a+c,b}(y,n)\;,
\;\;\;\hat{s}_{c_{\rm odd}ab}(y,n)=\hat{s}_{1,a+c-1,b}(y,n)\;,
\end{align}
or explicitly e.g.\
\begin{align}
\hat{s}_{000}(c,n)&=c+|n|\;,\;\;\;
\hat{s}_{010}(c,n)=\tfrac12\,[c^2-1+n^2+|n|c]\;,\\
\hat{s}_{020}(c,n)&=\tfrac16\,[3c(c^2-1+n^2)+|n|(3c^2+2n^2-2)]\;,\\
\hat{s}_{011}(c,n)&=\tfrac16\,[3c(c^2-1)+|n|(3c^2+n^2-4)]\;,\\
\hat{s}_{100}(c,n)&=-\tfrac{n}2\,[c+|n|]\;,\;\;\;
\hat{s}_{110}(c,n)=-\tfrac{n}6[3c^2+2n^2-2+|n|3c]\;.
\end{align}
Note that $g_{c,j}(a,b,d+1)/\Gamma(\alpha+\tfrac{d}2)$ 
in \eq\nr{eq:fPC3d}
needs a proper $\e\rightarrow0$ limit.
Meanwhile, the $j$ reduce to trigonometric functions, cf.\ \eq\nr{eq:jtrig},
for which the integral
in $h$ evaluates to exponentials times polynomials,
\begin{align}
h_{d,s}(x,y)&\approx
\frac{(-1)^{\frac{d-1}2}}{2\,e^{x+y}}\,\theta(x-y)
\( p^+_{d,s}(x,y)+ e^{2y}\,p^-_{d,s}(x,y) \)
+(x\leftrightarrow y)
+\order{\e}\;,
\end{align}
where the polynomials $p^\pm$ contain even and odd parts 
$t^\pm$ of reverse Bessel polynomials
\begin{align}
p^\pm_{d,s}(x,y) &=
\sum_{k=1}^s {s-1\choose k-1} t_{1,k+\frac{d-5}2}(x)\lk
y^{d-2}\,t_{0,s-k-\frac{d-1}2}(y) 
\pm y^{2s-2k}\,t_{0,k-s+\frac{d-3}2}(y)
\rk\;,\\
t_{b,n}(x)&=(2n-1)t_{b,n-1}(x)+x^2\,t_{b,n-2}(x)
\;,\quad
t_{b,0}(x)=1
\;,\quad
t_{b,1}(x)=1+bx\;,\\
\Rightarrow&\;t_{b,n<0}(x)=x^{2n+1}\lk
b\,t^+_{-n-1}(x)+t^-_{-n-1}(x) \rk
\;,\quad
t_{b,n\ge0}(x)=t^+_{n}(x)+b\,t^-_{n}(x)
\;,\nn
&\;t^\pm_n(x)=\sum_{k=0}^n
\frac{(2n-k)!}{(n-k)!k!}\,\frac{(2x)^k}{2^{n+1}}\(1\pm(-1)^k\)\;,
\end{align}
or explicitly e.g.\
\begin{align}
p^\pm_{3,3}(x,y) &=  
(y\pm1)(3 + 2 x)
\pm(x + x^2 + y^2)\;,\\
p^\pm_{5,3}(x,y) &=  
(y\pm1)(15 + 15 x + 6 x^2 + x^3 + y^2 + x y^2)
\pm\nn&
\pm y^2(5 + 5 x + 2 x^2 )\;,\\
p^\pm_{5,4}(x,y) &=  
(y\pm1)(105 + 105 x + 45 x^2 + 10 x^3 + x^4 + 10 y^2 + 10 x y^2 + 3 x^2 y^2)
\pm\nn&
\pm y^2(35 + 35 x + 15 x^2 + 3 x^3 + y^2 + x y^2)\;,\\
p^\pm_{7,3}(x,y) &=  
(y\pm1)(315 + 315 x + 135 x^2 + 30 x^3 + 3 x^4 + 30 y^2 + 30 x y^2 + 
 12 x^2 y^2 + 2 x^3 y^2)
\pm\nn&
\pm y^2(105 + 105 x + 45 x^2 + 10 x^3 + x^4 + 3 y^2 + 3 x y^2 + x^2 y^2)\;,
\end{align}
such that (using $\hat{f}(-n)=(-1)^c\,\hat{f}(n)$)
\begin{align}
\la{eq:D28}
\VfI(d,s_{1...8}) &=
\frac{T^2}{(4\pi)^{\frac{3d-1}2}}\,
\sum_{n=1}^\infty
\intx
\intxy\;
\frac{(4\pi)^{\frac{3d-1}2}}{(4\pi^2)^{d-1}}\,
\frac{(2\pi T)^{3d+1+s_{678}-2s_{12345}}\pi^{\frac{d}2-1}(-1)^{\frac{d-1}2}}
{2^{s_{12345}}\Gamma(s_1)\Gamma(s_2)\Gamma(s_3)\Gamma(s_4)\Gamma(s_5)
\Gamma(\frac{d}2)}\,
\times\nn&\times
\lk1+(-1)^{s_{678}}\rk
\frac{n^{s_6+4-d-2s_1}e^{-2nx}}{36(xy)^{2d-2s_{35}+s_8}}
\lk p^-_{d,s_1}(nx,ny)+e^{-2ny}p^+_{d,s_1}(nx,ny)\rk
\times\nn&\times
\lb\frac{\tilde{f}_{d,s_{247}}^{\Pi-B}(x,n)\tilde{f}_{d,s_{358}}^{\Pi-B}(y,n)}
{x^{2s_{35}-2s_{24}+s_7-s_8}}+(x\leftrightarrow y)\rb
+\order{\e}\;,\\
&\mbox{where~~}
\tilde{f}_{d,s_{247}}^{\Pi-B}(x,n)=
6\,x^{d+s_7-s_{24}}\,\hat{f}_{d,s_{247}}^{\Pi-B}(x,n)\;.
\end{align}

For our sum-integral $\intMcb$, we need to evaluate five specific cases
which follow from \eq\nr{eq:D28} and are all of similar structure,
such as e.g.\
($\csch^2(x)=\coth^2(x)-1$)
\begin{align}
 4 \VfI(3,31111,022)&=\frac{T^2}{(4\pi)^4}\,
\sum_{n=1}^\infty
\intx
\intxy\;
\frac{e^{-2nx}}{18 n^5 x^4 y^4}\,
\Big(3 + 3 n x + n^2 x^2 - n y - 2 n (1 + n x) y 
+\nn&+
 n^2 y^2 
- e^{-2 n y} \big(3 + 3 n x + n^2 x^2 + n y + 2 n (1 + n x) y 
+ n^2 y^2\big)\Big)
\times\nn&\times
\Big(3 + n x (3 + 3 n x - x^2) - 
 3 x^3 \big(n^2 \coth x + (n + \coth x) \csch^2 x\big)\Big)
\times\nn&\times
\Big(3 + n y (3 + 3 n y - y^2) - 
 3 y^3 \big(n^2 \coth y + (n + \coth y) \csch^2 y\big)\Big)
+\nn&+
\order{\e}\;.
\end{align}
All sums are of the form 
$s_a(z>0)=\sum_{n=1}^\infty e^{-2nz}/n^a={\rm Li}_a(e^{-2z})$
resulting in $s_{a>1}(z)={\rm Li}_a(e^{-2z})$, 
$s_1(z)=-\ln(1-e^{-2z})$ and $s_{a\le0}$ are 
polynomial of $\coth z$. For the numerical evaluation 
of the various $\VfI$ we use Mathematica \cite{mma} 
($\sum_n\rightarrow$ thousands of ${\rm Li}_j$; 
then we utilize
{\tt NIntegrate[...,\{x,0,1000\},\{y,0,x\},
MaxRecursion->100,WorkingPrecision->20]}).
As a result, weighting each piece by the prefactor
with it contributes to $\intMcb$, we obtain
\begin{align}
\la{eq:num1}
 4 \VfI(3,31111,022)&=\frac{T^2}{(4\pi)^4}   \lk 
+0.01854774(1)+\order{\e}\rk\;,\\
12 \VfI(5,31122,011)&=\frac{T^2}{(4\pi)^7}   \lk 
+0.02392697(1)+\order{\e}\rk\;,\\
 9 \VfI(5,41111,000)&=\frac{T^2}{(4\pi)^7}   \lk 
+0.00006691(1)+\order{\e}\rk\;,\\
 6 \VfI(5,32111,000)&=\frac{T^2}{(4\pi)^7}   \lk 
+0.00100101(1)+\order{\e}\rk\;,\\
\la{eq:num2}
60 \VfI(7,33311,000)&=\frac{T^2}{(4\pi)^{10}}\lk 
+0.01888983(1)+\order{\e}\rk\;.
\end{align}

%
\subsubsection{Second part $\VfII$ of \eq\nr{eq:Vf}}

The second part of $V^{\rm f}$ as in \eq\nr{eq:Vf}
is (this is only half of it; the other half is 
$247\leftrightarrow 358$)
\begin{align}\la{eq:VfIIh}
\VfIIh(d,s_{1...8}) &\equiv
\sumintp{P} \frac{(\Pz)^{s_6}}{(P^2)^{s_1}}
\(\Pi_{s_{247}}-\Pi_{s_{247}}^C\)
\(\Pi_{s_{358}}^B-\Pi_{s_{358}}^D\)\\
&=
\frac{(2\pi T)^{3d+3+s_{678}-2s_{12345}}}
{\pi^{\frac{3(d+1)}2}\,2^{2s_{12345}+2d}}\,
\frac{\Gamma(\frac12)}{\Gamma(\frac{d}2)}\,
\frac{1+(-1)^{s_{678}}}{\Gamma(s_2)\Gamma(s_4)}\,
\sum_{n=1}^\infty n^{d+3+s_{68}-2s_{135}}
\intx\,e^{-2nx}
\times\nn&\times
x^{s_{24}-d}\,
\hat{f}^{\Pi-C}_{d,s_{247}}(x,n)\,
\tilde{f}_{d,s_{1358}}(nx,\tfrac{\alpha}{4\pi^2 n^2},\e)\;,
\\
\tilde{f}_{d,s_{1358}}(x,y,\e) &= 
\sum_{j=0}^{[s_8/2]}2^j\frac{e^x}\pi
\intz\,
\frac{z\,j_{\frac{d}2-1}(xz)}{\(\frac{1+z^2}2\)^{s_{135}-\frac{d+1}2-j}}
\(1\!-\!\(\tfrac{1+z^2}{y}\)^\e\)
\frac1\e\;\e(4\pi)^{\frac{d+1}2}\,g_{s_8,j}(s_3,s_5,d\!+\!1)
\nn
&\approx
\ln\(\frac{xy e^\gammaEs}{2}\)\tilde\ell_{d,s_{1358}}(x)
+e^{2x}{\rm Ei}(-2x)\tilde\ell_{d,s_{1358}}(-x)
+\bar\ell_{d,s_{1358}}(x)
\!+\!\order{\e}\;,\\
{\tilde\ell\choose\bar\ell}_{d,s_{1358}}(x) &\equiv
\sum_{j=0}^{[s_8/2]} 2^j
\underbrace{\e(4\pi)^{\frac{d+1}2}\,g_{s_8,j}(s_3,s_5,d+1)}_{\mathbbm{Q}+\order{\e}}
\underbrace{{\ell_1\choose\ell_2}(d,s_{135}-\tfrac{d+1}2-j,x)}_{\mbox{poly(x)}}\;,\\
\ell_1(d,s,x)&\equiv
\frac{x^{d-2}}{\Gamma(s)}\,
\underbrace{e^x\,x^{s-\frac{d}2}\sqrt{\frac\pi2}\,
K_{\frac{d}2-s}(x)}_{\mbox{reverse Bessel poly}}
= \frac{x^{d-2}}{\Gamma(s)}\,\theta_{s-\frac{d+1}2}(x)\;,\\
&\quad 
\theta_{n<0}(x)=x^{1+2n}\theta_{-n-1}(x)\;,\quad
\theta_{n\ge0}(x)=\sum_{k=0}^n\frac{(2n-k)!}{(n-k)!k!}\,\frac{(2x)^k}{2^n}
\;,\\
\ell_2(d,s,x)&\equiv
\frac{e^x}\pi
\intz\,
\frac{z \sqrt{\frac\pi2}\,(xz)^{\frac{d}2-1}J_{\frac{d}2-1}(xz)}
{\(\frac{1+z^2}2\)^s}\,
\ln\frac1{1+z^2}\,
-\nn&-
\ln\(\frac{x e^\gammaEs}2\) \ell_1(d,s,x)
-e^{2x} {\rm Ei}(-2x) \ell_1(d,s,-x)\;.
\end{align}
\mbox{}From $\tilde\ell$ and $\bar\ell$, the cases that are useful for us read
\begin{align}
\bar{\ell}_{\{3,3112\},\{5,3121\},\{5,4110\},\{5,3210\},
\{5,3110\},\{7,3310\}}(x) &= \{2x^2,-2x^3,-x^3,3x^3,-2x^2,x^4\}/12\;,\nn
\bar{\ell}_{\{3,3112\},\{5,3121\},\{5,4110\},\{5,3210\},
\{5,3110\},\{7,3310\}}(x) &= -\{2x^2\!-\!2x,-2x^3,-x^3,3x^3,-4/3x^2,x^4\}/8\;.
\nonumber
\end{align}
The first part of $\hat{f}^{\Pi-C}=\hat{f}^{\Pi-B}-\hat{f}^{C-B}$
is defined in \eqs\nr{eq:fPCd} and \nr{eq:fPC3d} above, 
while the latter part is given by
\begin{align}
\hat{f}^{C-B}_{d,abc}(x,n) &=
\frac{1+(-1)^c}2\,\zeta(2a-c-d)\,
\frac{|n|^{\frac{d-1}2}x^{\frac{d+1}2}}
{2^{\frac{d-1}2}\pi^{\frac12}}\,
2^a\,\Gamma(a-\tfrac{d}2)\,\frac{\kappa_{b-\frac{d}2}(|n|x)}{|n|^b x^a}
+\\&+\nonumber
\lk-{\rm sign}(n)\rk^c
\sum_{n=0}^{[c/2]}{c\choose 2n}\zeta(2b-2n-d)|n|^{c-2n}
\frac{|n|^{\frac{d-1}2}x^{\frac{d+1}2}}
{2^{\frac{d-1}2}\pi^{\frac12}}\,
2^b\,\Gamma(b-\tfrac{d}2)\,\frac{\kappa_{a-\frac{d}2}(|n|x)}{|n|^a x^b}\;.
\end{align}
We can therefore write the finite parts 
$\VfII+\order{\e}=\VfIIh(d,s_{1...8})+\VfIIh(d,s_{13254687})+\order{\e}$
as
\begin{align}
 4 \VfII(3,31111,022)&=\frac{T^2}{(4\pi)^4}
\sum_{n=1}^\infty
\intx\,
\frac{2x}{3}\,
\hat{f}^{\Pi-C}_{3,112}(x,n)
\lk {\rm Ei}(-2nx)
+e^{-2nx}\ln\(\frac{\alpha_1 e^{\gamma-\frac32}}{16\pi^2}\,\frac{2x}{n}\)
+\frac{3\,e^{-2nx}}{2nx}
\rk\;,
\nn
12 \VfII(5,31122,011)&=\frac{T^2}{(4\pi)^7}
\sum_{n=1}^\infty
\intx\,
\frac{4x}{3}\,
\hat{f}^{\Pi-C}_{5,121}(x,n)
\lk {\rm Ei}(-2nx)
-e^{-2nx}\ln\(\frac{\alpha_2 e^{\gamma-\frac32}}{16\pi^2}\,\frac{2x}{n}\)
\rk\;,
\nn
 9 \VfII(5,41111,000)&=\frac{T^2}{(4\pi)^7}
\sum_{n=1}^\infty
\intx\,
\frac{1}{n}\,
\hat{f}^{\Pi-C}_{5,110}(x,n)
\lk {\rm Ei}(-2nx)
-e^{-2nx}\ln\(\frac{\alpha_3 e^{\gamma-\frac32}}{16\pi^2}\,\frac{2x}{n}\)
\rk\;,
\nn
 6 \VfII(5,32111,000)&=\frac{T^2}{(4\pi)^7}
\sum_{n=1}^\infty
\intx\,
\frac{-2}{3}\,
\hat{f}^{\Pi-C}_{5,210}(x,n)
\lk {\rm Ei}(-2nx)
+e^{-2nx}\ln\(\frac{\alpha_4 e^{\gamma-1}}{16\pi^2}\,\frac{2x}{n}\)
\rk
+\nn&+
\frac{T^2}{(4\pi)^7}
\sum_{n=1}^\infty
\intx\,
\frac{-1}{n}\,
\hat{f}^{\Pi-C}_{5,110}(x,n)
\lk {\rm Ei}(-2nx)
-e^{-2nx}\ln\(\frac{\alpha_3 e^{\gamma-\frac32}}{16\pi^2}\,\frac{2x}{n}\)
\rk\;,
\nn
60 \VfII(7,33311,000)&=\frac{T^2}{(4\pi)^{10}}
\sum_{n=1}^\infty
\intx\,
\frac{2x}{3}\,
\hat{f}^{\Pi-C}_{7,310}(x,n)
\lk {\rm Ei}(-2nx)
+e^{-2nx}\ln\(\frac{\alpha_5 e^{\gamma-\frac32}}{16\pi^2}\,\frac{2x}{n}\)
\rk\;.
\nonumber
\end{align}
Note that line 3 and 5 cancel exactly (assuming they are finite).
The reason is that $\Pi^B_{210}=\frac{g_{00}(2,1,d+1)}{[P^2]^{3-\frac{d+1}2}}
=\frac{3-d-1}{P^2}\,\frac{g_{00}(1,1,d+1)}{[P^2]^{2-\frac{d+1}2}}
=\frac{2-d}{P^2}\,\Pi^B_{110}$ such that
the relevant contributions to $\intMcb$ combine with a pre-factor of 
$\order{\e}$:
\begin{align}
\intMcb^{\rm nz,f} &\ni 2d{\cal D}^+\lb 3\VfIIh(41111,000,\alpha_3)
+\VfIIh(31211,000,\alpha_3)\rb\nn
&=2d{\cal D}^+\lb 3\VfIIh(41111,000,\alpha_3)
+(2-d)\VfIIh(41111,000,\alpha_3)\rb\nn
&=2d{\cal D}^+\,\Big\{(5-d)\VfIIh(41111,000,\alpha_3)\Big\}\nn
&=2d(3-d){\cal D}^+\VfIIh(41111,000,\alpha_3)\;.
\end{align}
Thus, also the contributions from $V^{\rm d}$ should not contain 
$\alpha_3$ up to their constant terms, which is indeed
the case, serving as a small check of our expressions.
Furthermore, it appears convenient to choose 
$\alpha_1=\alpha_2=\alpha_5=16\pi^2\,e^{3/2-\gammaEs}$
and $\alpha_4=16\pi^2\,e^{1-\gammaEs}$, although the $\gamma$
could also remain in the log, see \eq\nr{eq:c3check}.

The various functions $\hat{f}^{\Pi-C}$ above are 
(omitting the arguments $(x,n)$ on the lhs)
\begin{align*}
\hat{f}^{\Pi-C}_{3,112}&=\frac1{30x^3}\,
\Big(x^4 - 15 + 5 n x (x^2 - 3) - 5 n^2 x^2 (3 + x^2) + 
 15 x^3 \big(n^2 \coth x + (n + \coth x) \csch^2 x\big)\Big)\;,\\
\hat{f}^{\Pi-C}_{5,121}&=\frac{n}{6x^2}\,
\Big(6 + x \big(3 x + n (3 + x^2)\big) - 3 x \coth x (1 + n x + x \coth x)\Big)\;,\\
\hat{f}^{\Pi-C}_{5,110}&=-\frac1{90x^3}\,
\Big(225 + 135 n x + 15 n x^3 + x^4 + n x^5 
-\nn&- 
45 x \big(2 (1 + n x) \coth x + x (2 + n x + x \coth x) \csch^2 x\big)\Big)\;,\\
\hat{f}^{\Pi-C}_{5,210}&=-\frac1{180x^2}\,
\Big(270 + 120 x^2 + x^4 + 30 n x (3 + x^2)-90x\coth x \big(2 + n x + x \coth x\big)\Big)\;,\\
\hat{f}^{\Pi-C}_{7,310}&=-\frac1{1890x^3}\,
\Big(9450 + x \big(3780 x + x^5 + 315 n^2 x (3 + x^2) + 315 n (12 + 5 x^2)\big) 
-\nn&-
 945 x \coth x \big(6 + 3 n x + (-1 + n^2) x^2 
+ x \coth x (3 + n x + x \coth x)\big)\Big)\;.
\end{align*}

A quick numerical evaluation
(summing up to $n_{\rm max}=10000$, and using Mathematica's
{\tt NIntegrate[...,,MaxRecursion->100,WorkingPrecision->60,AccuracyGoal->30]},
while estimating the error by fitting values
$n\in[10000,60000]$ and extrapolating to infinity)
of the (weighted) $\VfII$ parts then produces
(choosing all $\alpha_i=16\pi^2\,e^{3/2-\gammaEs}$)
\begin{align}
\la{eq:num3}
 4 \VfII(3,31111,022)&=\frac{T^2}{(4\pi)^4}   \lk 
+0.00775440(1)+\order{\e}\rk\;,\\
12 \VfII(5,31122,011)&=\frac{T^2}{(4\pi)^7}   \lk 
-0.00354681(1)+\order{\e}\rk\;,\\
 9 \VfII(5,41111,000)+
 6 \VfII(5,32111,000)&=\frac{T^2}{(4\pi)^7}   \lk 
+0.00295006(1)+\order{\e}\rk\;,\\
\la{eq:num4}
60 \VfII(7,33311,000)&=\frac{T^2}{(4\pi)^{10}}\lk 
-0.00503877(1)+\order{\e}\rk\;.
\end{align}
It might be possible to evaluate (some of) the $\VfIIh$ analytically.
We have not put further effort into this, as the numerical values given
above are fully sufficient for our purposes. 
Before leaving, let us record some integrals and some sums 
that might be useful in that respect:
\begin{align}
&\intz\,e^{-z}\,z^n
= \Gamma(n+1)\;,
\qquad\intz\,{\rm Ei}(-z)\,z^n
= \frac{\Gamma(n+1)}{n+1}\;,\quad{\rm Re}(n)>-1\;,\\
&\intz\,e^{-z}\,z^n\,(\gamma+\ln z)
= \Gamma(n+1)\,{\rm HarmonicNumber}(n)\;,\quad{\rm Re}(n)>-1\;,\\
&\intz\lk e^{-z}(\gamma+\ln z)-{\rm Ei}(-z)\rk
\frac1z
= \zeta(2)=\frac{\pi^2}6\;,\\
&\sum_{n=1}^\infty \frac{\ln n}{n^a} = -\zeta'(a)\;,
\quad
\sum_{n=1}^\infty \frac{e^{-2nx}}{n^a} = {\rm Li}_a(e^{-2x})\;,
\quad
\sum_{n=1}^\infty \frac{e^{-2nx}\ln n}{n^a} = 
-\partial_a {\rm Li}_a(e^{-2x})\;.
\end{align}
One can get some analytic parts from looking 
at the piece 
$\sim\hat{f}^{C-B}_{5,110}(x,n)=\frac{x}{90}\,(1+|n|x)$ of
\begin{align}
\la{eq:c3check}
\VfIIh(5,41111,000,\alpha_3) &\ni
\frac{T^2}{(4\pi)^7}
\sum_{n=1}^\infty
\intx\,
\frac{\hat{f}^{C-B}_{5,110}(x,n)}{18n}\lk 
e^{-2nx}\ln\(\frac{\alpha_3 e^{\gamma-\frac32}}{16\pi^2}\,\frac{2x}{n}\)
-{\rm Ei}(-2nx)
\rk \nn
&=\frac{T^2}{(4\pi)^7}\,
\frac{1}{18}\,
\sum_{n=1}^\infty
\bigg\{\frac1{108n^3}\underbrace{\frac3{20}
\intz\,
z(2+z)\lk e^{-z}\ln(z\,e^\gammaEs)-{\rm Ei}(-z)\rk}_{=1}
-\nn&-
\frac{\ln n}{90n^3}\underbrace{\frac14
\intz\,
z(2+z)e^{-z}}_{=1}\bigg\}
=\frac{T^2}{(4\pi)^7}\,
\frac{1}{18^2}\(\frac{\zeta(3)}6+\frac{\zeta'(3)}5\)\;,
\end{align}
where in the second line we have chosen $\alpha_3=16\pi^2e^{3/2}$, 
and which is confirmed by the corresponding analytic 
expression for the $(\Pi^C-\Pi^B)$ piece of $\hat{V}^{\rm d}$,
which can be extracted by considering only the $I\cdot I$ parts 
of \eq\nr{eq:fexpr}.

%
\subsubsection{Summing up}

Summing up \eqs\nr{eq:num1}-\nr{eq:num2} 
and \eqs\nr{eq:num3}-\nr{eq:num4},
we obtain the numerical coefficient 
of \eq\nr{eq:Mcbfnz} as
\begin{align}
n_1 &\approx +0.0645513(1) \;.
\end{align}

%
\subsection{Contribution to $\intMcb$ from finite parts  $V^{\rm z,f}$ 
of zero-modes}
\la{se:Zfin}

Here, we discuss the evaluation of the finite terms
given in \eq\nr{eq:Mfz} of the main text.

Working again in coordinate space, we need the Fourier transform for 
$\Pi^A$ of \eq\nr{eq:PiA}.
Being the sum of two terms, its first part can be read off from the 
last line of \eq\nr{eq:fPC3d}, while in the notation of
\eq\nr{eq:FTs} with $\Pz\!=\!0$ 
its second part (let us label it $E$ here) reads
\begin{align}
\la{eq:PiEdef}
\hat{f}_{d,ab0}^E(2\pi Tr,0) &=
\frac{\Gamma(d/2-a)\Gamma(d/2-b)}{\Gamma^2(1/2)}\,
\(\frac{2\pi T r}{2}\)^{a+b+1-d}\;,
\end{align}
whereas below we also need the special case 
$\hat{f}_{3,222}^E(x,0)=1$.

What we need to evaluate is the first line of \eq\nr{eq:zeroModeDec}
for which, using \eq\nr{eq:jtrig} we obtain
\begin{align}
\la{eq:Sfint}
S^{\rm f}(d;s_{1...5};s_7,s_8) &=
\sumint{P} \frac{\delta_{\pz}}{[P^2]^{s_1}}\,
\(\Pi_{s_{247}}-\Pi_{s_{247}}^A\)
\(\Pi_{s_{358}}-\Pi_{s_{358}}^A\)\\
&=\frac{(2\pi T)^{3d+3+s_{78}-2s_{12345}}2^{1-2d-s_{12345}}}
{\Gamma(s_1)\Gamma(s_2)\Gamma(s_3)\Gamma(s_4)\Gamma(s_5)
\Gamma(d/2)\pi^{1+3d/2}}\,
\intx\intxy\;\hat{h}_{d,s_1}(x,y)
\times\nn&\times
\Big[x^{s_{24}-d}\,y^{s_{35}-d}\,
\hat{f}_{s_{247}}^{\Pi-A}(x,0)\,\hat{f}_{s_{358}}^{\Pi-A}(y,0)
+\{x\!\leftrightarrow\!y\}\Big]\;,\\
\mbox{with~~}\hat{h}_{d,s}(x,y)&=
\frac{\Gamma(d/2-s)y^{d-2}}{2^s\,\Gamma(d/2)}\,
x^{2s-2}\,{}_2F_1(d/2-s,1-s,d/2,y^2/x^2)
\;\;\;\;(d\!>\!2s\!>\!0)\;,\nn
\mbox{and special cases}&\;\;\;\hat{h}_{\{31,51,52,72\}}(x,y)=
\big\{y,\tfrac{y^3}3,\tfrac{y^3}3\,\tfrac{5x^2+y^2}5,
\tfrac{y^5}{15}\,\tfrac{7x^2-3y^2}7\big\}\;.
\end{align}
The function $\hat{h}$ in fact originates from
\begin{align*}
\frac{2^s\,\Gamma(s)}{\pi}\,\intz\,z^{3-d-2s}\,
j_{d/2-1}(x\,z)\,j_{d/2-1}(y\,z) &=
\theta(x-y)\,\hat{h}_{d,s}(x,y) + \theta(y-x)\,\hat{h}_{d,s}(y,x)\;,
\end{align*}
which corresponds to the overall $\int_{\vec p}$ of \eq\nr{eq:Sfint}
after performing the angular integrals.

To get an explicit expression of $f^{\Pi-A}$, 
we simply refer to \eq\nr{eq:fPC3d},
taken at $n=0$ and where we only have to insert the theta function
of \eq\nr{eq:PiA} into the second line, 
as well as \eq\nr{eq:PiEdef} above, 
again multiplied by the theta function.
For the specific cases that we are interested in,
we get the compact expressions
\begin{align}
\tfrac45\,S^{\rm f}_{0110}(3,12121,22) 
&=\frac{T^2}{(4\pi)^4}\,\intx\intxy\,
\frac1{5 x^4 y^3}\,\Big[
f_1(x,y)+\{x\!\leftrightarrow\!y\}
\Big]
\;,\\
f_1(x,y)&=
x^5 (\coth x - 1) \big(y^3 \coth(y) \csch^2 y - 1\big)\;,\nn
\tfrac45\,S^{\rm f}_{1010}(3,12211,22)
&=\frac{T^2}{(4\pi)^4}\,\intx\intxy\,\frac1{5x^2y}\,
\big(1-x^2\csch^2x\big)\big(1-y^2\csch^2y\big)\\
&=\frac{T^2}{(4\pi)^4}\,\intx\,\frac1{5x^2}\,
\big(1-x^2\csch^2x\big)\big(\ln(x\,\csch x)+x\coth x-1\big)
\;,\nn
-\tfrac65\,S^{\rm f}_{0110}(5,22121,00)
&=\frac{T^2}{(4\pi)^7}\,\intx\intxy\,
\frac{-2(5x^2+y^2)}{75 x^6 y^3}\,\Big[
f_3(x,y)+\{x\!\leftrightarrow\!y\}
\Big]
\;,\\
f_3(x,y)&=
x^5 (\coth x - 1) \big(2 y \coth y - 5 + y^2 (2 + y \coth y) \csch^2 y\big)\;,\nn
-\tfrac{12}5S^{\rm f}_{0110}(5,12221,00)
&=\frac{T^2}{(4\pi)^7}\,\intx\intxy\,
\frac{-4}{15 x^4 y}\,\Big[
f_4(x,y)+\{x\!\leftrightarrow\!y\}
\Big]
\;,\\
f_4(x,y)&=
x^3 (\coth x - 1) \big(-3 + y \,\csch^2(y)\, (y + \sinh(2 y))\big)\;,\nn
96\,S^{\rm f}_{0110}(7,23321,00)
&=\frac{T^2}{(4\pi)^{10}}\,\intx\intxy\,
\frac{4(7x^2-3y^2)}{525 x^6 y}\,\Big[
f_5(x,y)+\{x\!\leftrightarrow\!y\}
\Big]
\;,\\
f_5(x,y)&=
x^3 (\coth x \!-\! 1) (2 \!+\! x \!+\! x \coth x) \big(6 y \coth y \!-\! 10 \!+\! 
   y^2 (3 \!+\! y \coth y) \csch^2 y\big)\;,\nn
84\,S^{\rm f}_{0110}(7,23222,00)
&=\frac{T^2}{(4\pi)^{10}}\,\intx\intxy\,
\frac{7x^2-3y^2}{75 x^6 y}\,\Big[
f_6(x,y)+\{x\!\leftrightarrow\!y\}
\Big]
\;,\\
f_6(x,y)&=
x^3 (\coth x \!-\! 1) (2 \!+\! x \!+\! x \coth x) \big(2 y \coth y \!-\! 5 \!+\! 
   y^2 (2 \!+\! y \coth y) \csch^2 y\big)\;.\nonumber
\end{align}
The $y$\/-integration can be explicitly performed in most 
(all but the last two) cases,
giving a result containing zetas and (poly)logarithms,
whereas the remaining $x$\/-integration can be approximated
numerically, to give 
\begin{align}
\la{eq:zres1}
\tfrac45\,S^{\rm f}(3,12121,22) &\approx \tfrac{T^2}{(4\pi)^4}\big[
-0.01005114745(1)+\order{\e}\big]\;,\\
\tfrac45\,S^{\rm f}(3,12211,22) &\approx \tfrac{T^2}{(4\pi)^4}\big[
+0.15213227462(1)+\order{\e}\big]\;,\\
-\tfrac65\,S^{\rm f}(5,22121,00) &\approx \tfrac{T^2}{(4\pi)^7}\big[
-0.00891125885(1)+\order{\e}\big]\;,\\
-\tfrac{12}5S^{\rm f}(5,12221,00) &\approx \tfrac{T^2}{(4\pi)^7}\big[
-0.10673211253(1)+\order{\e}\big]\;,\\
96\,S^{\rm f}(7,23321,00) &\approx \tfrac{T^2}{(4\pi)^{10}}\big[
+0.20021689747(1)+\order{\e}\big]\;,\\
\la{eq:zres6}
84\,S^{\rm f}(7,23222,00) &\approx \tfrac{T^2}{(4\pi)^{10}}\big[
+0.02318282360(1)+\order{\e}\big]\;.
\end{align}

%
\subsubsection{Summing up}

Summing up \eqs\nr{eq:zres1}-\nr{eq:zres6}
we obtain the numerical coefficient 
of \eq\nr{eq:Mfz} as
\begin{align}
n_2 &\approx +0.24983747686(1) \;.
\end{align}

%
\section{Cross-checks}
\la{se:checks}

Using the generic formulae for spectacles-type
sum-integrals as derived in the main text,
we can check two particular cases, $\intV$ and $\intVb$,
against the previously known results from the literature.

First, let us check $\intV$ of \se2 in \cite{Schroder:2012hm}:
In the language developed here,
\begin{align*}
\intV &= V(11111,000) = V^{\rm f}+V^{\rm d}+V^{\rm z}
\;=\; {\cal B}+2{\cal C}+V^{\rm d}(11111,000)+{\cal A}+S^{\rm d}(11111,00)\;,
\end{align*}
with $\{{\cal A},{\cal B},{\cal C}\}$ as defined in \cite{Schroder:2012hm}
and whose expansion (using $p^\pm_{31}(x,y)=\pm1$ 
as well as \eqs\nr{eq:VdRes}, \nr{eq:Sd} and $L(111,00,d)=0$) 
exactly reproduces \eqs(2.15)-(2.17) of Ref.~\cite{Schroder:2012hm}.

Second, let us check $\intVb$ of \cite{Ghisoiu:2012kn}:
In the language developed here,
\begin{align*}
\intVb &= V(21111,002) = V^{\rm f}+V^{\rm d}+V^{\rm z}
\;=\; \cS_2+\cS_4+\cS_6+V^{\rm d}(21111,002)+S(21111,02)\;,
\end{align*}
where IBP on $S(21111,02)$ gives the reduction
\begin{align*}
\cS_1 &= S(21111,02)
\;=\;\frac1{d-6}\lb
S(12111,02)+S(12111,20)-I_2^0\,A(211,2)-I_2^2\,A(211,0)
\rb
\end{align*}
for which
\begin{align}
S(12111,02) &= \cS_{1a}+S^{\rm d}(12111,02)\;, \\
S(12111,20) &= \cS_{1b}+S^{\rm d}(12111,20)\;.
\end{align}
Collecting and expanding (using $p^\pm_{32}(x,y)=y\pm(x+1)$;
\eqs\nr{eq:VdRes}, \nr{eq:Sd}; 
as well as $L(111,00,d)=0$; $L(211,20,d)=0$; 
and $L(211,02,d)$ from \eq\nr{eq:L21102}), 
one reproduces \eqs(4.1)-(4.2) of Ref.~\cite{Ghisoiu:2012kn}.


\end{appendix}



\end{document}